\documentclass[aps,prb,twocolumn,showpacs,10pt,superscriptaddress,floatfix,longbibliography]{revtex4-2}

\usepackage{newtxtext}
\usepackage{newtxmath}
\usepackage{graphicx}
\usepackage{subfigure}
\usepackage{amsmath}
\usepackage{amsfonts}
\usepackage{mathrsfs}
\usepackage{float}
\usepackage{booktabs}
\usepackage{bbm}
\usepackage{xcolor}
\usepackage[colorlinks=true]{hyperref}
\usepackage{ulem}
\usepackage[T1]{fontenc}
\usepackage{svg}

\usepackage{appendix}

\begin{document}

\title{Semiclassical theory for proximity-induced superconducting systems with spin-orbit coupling}

\author{Zhen-Cheng Liao}
\affiliation{School of Physics, Sun Yat-sen University, Guangzhou 510275, China}

\author{Cong Xiao}
\email{congxiao@fudan.edu.cn}
\affiliation{Interdisciplinary Center for Theoretical Physics and Information Sciences (ICTPIS), Fudan University, Shanghai 200433, China}

\author{Zhi Wang}
\email{wangzh356@mail.sysu.edu.cn}
\affiliation{School of Physics, Sun Yat-sen University, Guangzhou 510275, China}
\affiliation{Guangdong Provincial Key Laboratory of Magnetoelectric Physics and Devices, Sun Yat-sen University, Guangzhou 510275, China}

\author{Qian Niu}
\affiliation{School of Physics, University of Science and Technology of China, Hefei, Anhui 230026, China }
\begin{abstract}
We develop a semiclassical theory of superconducting quasiparticles for proximity-induced superconducting systems, where spin-orbit coupling plays a critical role in shaping the quasiparticle dynamics. We reveal the structure of superconducting Berry curvatures in such systems, and derived the superconducting Berry curvature induced thermal Edelstein effect and spin Nernst effect. We calculate these two thermo-spin responses with model systems where Rashba spin-orbit coupling, proximity induced superconductivity, and ferromagnetic order are coexisting. 
\end{abstract}

\maketitle
\section{Introduction}
Berry curvature effects are widely present in electron systems that break time-reversal or inversion symmetry \cite{xiao2010berry}, inducing a variety of important response phenomena \cite{nagaosa2010anomalous,smejkal2022anomalous,chang2023colloquium}. The Berry curvature effects have also received increasing interests in the context of superconducting quasiparticles \cite{cvetkovic2015berry,murray2015majorana,liang2017wavepacket,wang2021berry,zhou2023topological,zhang2023topological,yang2024optical,Hsu2025}. In particular, the superconducting Berry curvatures 
defined in the phase space spanning momentum and position dictate the anomalous thermal Hall transport and the density of states modulation for superconducting quasiparticles \cite{cvetkovic2015berry,murray2015majorana,wang2021berry,Hsu2025}.

One important source of Berry curvature in Bloch band is the spin-orbit coupling (SOC) \cite{manchon2015new,ren2016topological}, and one expects the same for superconducting quasiparticles. For intrinsic superconductors, the SOC may mix the singlet-triplet components of the Cooper pair \cite{gorkov2001superconducting,smidman2017superconductivity,fischer2023}. This drastic change in condensate should naturally influence the Berry curvature of quasiparticles. For proximity induced superconducting systems, the SOC can qualitatively change the topology of the superconductors, transforming a trivial $s$-wave superconductor with Chern number zero to an effective chiral $p$-wave superconductor with Chern number one \cite{fu2008superconducting,zhang2008p+ip,sato2009nonabelian,sau2010generic,liu2013d+id,qi2011topological,sato2017topological}. Since the Chern number is simply the integral of the momentum-space Berry curvature \cite{read2000paired,frolov2020topological,flensberg2021engineered}, the Berry curvature must experience a drastic change during this topological transformation due to SOC. However, the properties of such Berry curvature and their possible role in nonequilibrium spintronic responses of proximity-induced superconducting systems still remain elusive, leaving many basic questions not clarified. For instance, how does the interplay of pairing and band physics shape the structure of superconducting Berry curvatures around the electron Fermi surface? How will such Berry curvature induce the thermal Edelstein effect \cite{freimuth2014DMI-SOT,xiao2016spin,freimuth2016inverse,Shitade2019,dong2020berry} and spin Nernst effect \cite{Bose2018SNE,Bose2018SNT,kim2020SNT,zhang2020SNE,jain2023thermally}, which measure respectively the generation of spin density and spin current by temperature gradient? To address these questions will be a significant expansion of superconducting spintronics \cite{Linder2015superconducting,Han2020spin,Yang2021boosting}. 

In this work, we address the above questions by constructing the theory of superconducting Berry curvatures in proximity-induced superconducting systems with SOC. By a semiclassical wavepacket approach, we derive the Berry curvature induced thermal Edelstein effect and spin Nernst effect in such proximity superconductors.
We implement the theory to study a two-dimensional superconducting toy model where Rashba SOC, magnetic order, and $s$-wave superconductivity are coexisting due to the proximity effect. We show that the Berry curvature induced thermo-spin responses exist in a wide range of parameter space. Our theory lays the groundwork for superconducting Berry curvature physics in proximity superconductors with SOC.


This paper is organized as follows. In Sec. \ref{sec:semi-methods}, we establish the semiclassical methods for studying the proximity-induced superconducting systems with SOC. In Sec. \ref{sec:response}, we propose two thermo-spin responses in superconductors using the semiclassical methods. In Sec. \ref{sec:curvature}, we systematically analyze the superconducting Berry curvatures that dominate the thermo-spin responses. Finally, the numerical results for a minimal model are presented in Sec. \ref{sec:model}. 

\section{Semiclassical approach}\label{sec:semi-methods}


In this section, we extend the framework of the semiclassical approach for superconducting quasiparticles to include the proximity-induced superconducting systems with SOC. 
A proximity-induced spin-singlet superconducting gap is assumed to be present for the superconductor. With this assumption, we employ a Bogoliubov-de-Gennes (BdG) Hamiltonian to describe the system and construct the wavepacket on one BdG band. Then we follow the motion of the wavepacket center in both real and momentum space, and derive the equations of motion. In the equations of motion, we reveal various Berry curvature effects.

\subsection{Local Hamiltonian}
For proximity-induced superconducting systems, the pairing symmetry of the proximity-induced superconducting gap is well-defined by a spin-singlet $s$-wave or $d$-wave pairing. The SOC system can be described by a mean-field Hamiltonian
\begin{align}\label{eq:Hamiltonian}
\hat{H}&=\int d{\bf r} c_{\sigma}^{\dagger}({\bf r})h_{\sigma \sigma'}({\bf r},-i\nabla-{\bf A}({\bf r});\beta({\bf r}))c_{\sigma'}({\bf r})\nonumber\\
&+\iint d{\bf r}_1 d{\bf r}_2[\Delta_{\sigma\sigma'}({\bf r}_1,{\bf r}_2)c_{\sigma}^{\dagger}({\bf r}_1)c_{\sigma'}^{\dagger}({\bf r}_2)+h.c.],
\end{align}
where $\int d{\bf r}$ denotes the integration over the real space, $c_{\sigma}^{\dagger}({\bf r})$ is the electron creation operator that creates an electron with spin $\sigma=\{\uparrow,\downarrow\}$ at position ${\bf r}$, $h_{\sigma \sigma'}({\bf r},-i\nabla-{\bf A}({\bf r});\beta({\bf r}))$ is the electron Hamiltonian incorporating the SOC, the gauge field ${\bf A}({\bf r})$, and various external perturbations from impurities, the boundary, or other external fields represented with $\beta({\bf r})$, $\Delta_{\sigma\sigma'}({\bf r}_1,{\bf r}_2)$ denotes the proximity-induced superconducting gap with ${\bf r}_{1}$ and ${\bf r}_{2}$ being the positions of the two electrons in a Cooper pair. Here and after, we take the Einstein summation convention for $\sigma$ and $\sigma'$,  and set $e=1$ and $\hbar=1$.

The external perturbations break the translational invariance of the crystal, hindering the application of the band theory. The semiclassical approach deals with this difficulty by studying the dynamics of a wavepacket. In the limit that the spatial variation of the perturbation fields is smooth in comparison with the size of the wavepacket, the dynamics of each wavepacket is dominated by a Hamiltonian that locally preserves the translational invariance, namely the local Hamiltonian $\hat{H}_{c}$, in which all the external fields take values at the wavepacket center ${\bf r}_c$. The local electron Hamiltonian is  written straightforwardly as \cite{sundaram1999wavepacket}
\begin{equation} \label{eq:localelectron}
\hat{H}^0_{c} = \int d{\bf r}c_{\sigma}^{\dagger}({\bf r})h^c_{\sigma\sigma'}({\bf r},-i\nabla-{\bf A}({\bf r}_c);\beta({\bf r}_{c}))c_{\sigma'}({\bf r}),
\end{equation}
where the position operator $\mathbf{r}$ in $\mathbf{A}$ and $\beta$ is changed to $\mathbf{r}_c$.
The local approximation for the pairing Hamiltonian is more complicated due to the self-consistent constraint from the $U(1)$ gauge symmetry breaking of the superconductors. It requires that the gradient of the phase of the gap function with respect to ${\bf r}_c$ should equal to the finite momentum of the center of mass of the Cooper pair with a position $({\bf r}_1 + {\bf r}_2)/2$. Hence, the local pairing Hamiltonian should be written as
\begin{equation} \label{eq:localgap}
\hat{H}^{1}_{c} = \iint d{\bf r}_1 d{\bf r}_2[ \Delta^c_{\sigma\sigma'}c_{\sigma}^{\dagger}({\bf r}_1)c_{\sigma'}^{\dagger}({\bf r}_2)+h.c.],
\end{equation}
where the local gap function is written as
\begin{equation}
    \Delta^c_{\sigma\sigma'}=\Delta({\bf r}_c) \chi_{\sigma\sigma'}({\bf r}_1 - {\bf r}_2)e^{\frac{i}{2} ({\bf r}_1 + {\bf r}_2) \cdot{\bf \nabla_{{\bf r}_c}\varphi({\bf r}_{\mathbf{c}})}}.
\end{equation}
Here $\Delta({\bf r}_c)$ denotes the gap function at the center of the wavepacket, $\varphi({\bf r}_{\mathit{c}})$ is the phase of $\Delta({\bf r}_c) $, and $\chi_{\sigma\sigma'}({\bf r}_1 - {\bf r}_2)$ represents the relative wave function of two electrons in a Cooper pair. It is noted that an additional phase of $\frac{1}{2}({\bf r}_1+{\bf r}_2) \cdot \nabla \varphi({\bf r}_{c})$ is induced to satisfy the global gauge invariance of the local Hamiltonian, which mathematically requires that the vector potential ${\bf A}({\bf r}_c)$ and the gradient of the superconductor phase $\nabla_{{\bf r}_c} \varphi({\bf r}_c)$ are related through a pure gauge.

We then transform the local Hamiltonian to the band representation of the electron Hamiltonian $\hat{H}^0_c$. The transformation for the electron annihilation operator reads
\begin{equation}\label{eq:Fourier}
   c_{\sigma}({\bf r}) = \sum_{n{\bf k}} c_{n \bf{k}}\phi_{n\mathbf{k},\sigma}({\bf r})e^{i{\bf k}\cdot{\bf r}}e^{i{\bf A}(\mathbf{r}_{c})\cdot{\bf r}},
\end{equation}
where $\phi_{n\mathbf{k},\sigma}({\bf r})$ is the $\sigma$ component of the cell-periodic spinor Bloch function with $n$ being the band index and $\bf k$ being the crystal momentum. Using this transformation, the local Hamiltonian can be transformed to the band representation (see Appendix~\ref{app:local pairing} for details):
\begin{equation}
{\hat H}_{c}=\sum_{n\mathbf{k}} \xi^c_{n {\bf k}} c_{n\mathbf{k}}^{\dagger}c_{n\mathbf{k}} +\sum_{nn'{\bf k}}(\tilde{\Delta}_{nn'{\bf k}} c_{n\mathbf{k}+\frac{\bf q}{2}}^{\dagger}c_{n'-\mathbf{k}+\frac{\bf q}{2}}^{\dagger}+h.c.),
\label{eq:localelectron-momentum2}
\end{equation}
where $\xi^c_{n{\bf k}}$ is the eigen-energy of ${\hat h}^c$ measured from the chemical potential $\mu$, and we find the effective pairing gap in the band representation written as
\begin{equation}\label{eq:gapfunction}
     \tilde{\Delta}_{nn'{\bf k}}
= \Delta_{{\bf k}}  \int_{{\bf R},{{\bf r}}} \phi_{n\mathbf{k}+\frac{\bf q}{2},\sigma}^{*}({\bf R}+\frac{\bf r}{2}) \chi^c_{\sigma \sigma'}({\bf r})
\phi_{n'-\mathbf{k}+\frac{\bf q}{2},\sigma'}^{*}({\bf R}-\frac{\bf r}{2}), 
\end{equation}
where ${\bf q}={\bf \nabla\varphi}({\bf r}_{c})-2{\bf A}({\bf r}_{c})$ measures the finite momentum of the Cooper pair due to supercurrent velocity, ${\bf R} = ({\bf r}_1 + {\bf r}_2)/2$ and ${\bf r} = {\bf r}_1 - {\bf r}_2$ respectively represent the center of mass coordinate and the relative coordinate for the two electrons in a Cooper pair, $\int_{{\bf R},{{\bf r}}}$ represents the integration over one unit cell, $\chi^c_{\sigma \sigma'}({\bf r})$ is the cell periodic part of $\chi_{\sigma \sigma'}({\bf r})$, ${\Delta}_{\bf k}
= \Delta({\bf r}_c)\int d{\bf r} \chi^{e}({\bf r}) e^{i{\bf k}\cdot {\bf r} }$ with $\chi^{e}({\bf r})$ the envelop function of $\chi_{\sigma \sigma'}({\bf r})$. The ${\Delta}_{\bf k}$ manifest pairing symmetry of the proximity-induced gap. For example, $\chi^{e}({\bf r})=e^{-r/r_{0}}$ and $\chi^{e}({\bf r})=(x^{2}-y^{2})e^{-r/r_{0}}$ with $r_0$ a constant, give the conventional $s$-wave pairing gap and a typical $d$-wave pairing gap function, receptively. 

Now we introduce the Bogoliubov-de Gennes (BdG) Hamiltonian by defining the Nambu operators for Bloch electrons ${\tilde c}_{\mathbf{k}}^{\dagger}=(c_{1,{\bf k}+\frac{\bf q}{2}}^{\dagger},\cdots,c_{N,{\bf k}+\frac{\bf q}{2}}^{\dagger},c_{1,\mathbf{-k}+\frac{\bf q}{2}},\cdots,c_{N,\mathbf{-k}+\frac{\bf q}{2}})$ with $N$ being the number of relevant electron Bloch bands.
The local Hamiltonian is written as
\begin{equation}
\hat H_c=\frac{1}{2}\sum_{\mathbf{k}}\tilde{c}_{\mathbf{k}}^{\dagger} {H}_{c} ({\bf k})\tilde{c}_{\mathbf{k}},
\end{equation}
where the BdG matrix is given by
\begin{equation}\label{eq:BdG}
{H}_{c} ({\bf k})=\left(\begin{array}{cc}
{h}_{nn'}({\bf k}+\frac{\bf q}{2})& \tilde{\Delta}_{nn'}({\bf k})\\
\tilde{\Delta}^\dagger_{nn'}({\bf k})  & -{h}^*_{nn'}(-{\bf k}+\frac{\bf q}{2}) 
\end{array}\right).
\end{equation}
The diagonal block of the BdG matrix is the spectrum of the electron $h_{nn'{\bf k}}=\xi^c_{n{\bf k}}\delta_{nn'}$. The off-diagonal Block is the effective pairing gap in the band representation \eqref{eq:gapfunction}. 

We can solve the eigenvalue problem for this BdG matrix to obtain the Bogoliubov quasiparticle wavefunction
\begin{equation}
{H}_{c} ({\bf k})
\left(\begin{array}{c}U_{\mathbf{k}}\\
V_{\mathbf{k}}\end{array}\right)=E({\bf k})\left(\begin{array}{c}
U_{\mathbf{k}}\\
V_{\mathbf{k}}
\end{array}\right),\label{eq:Bogoliubov}
\end{equation}
where $E({\bf k})$ is the BdG spectra, $U_{\mathbf{k}}=(U_{1{\bf k}},\dots,U_{N{\bf k}})^T$ denotes the amplitude of the electron component of the eigenvector, and $V_{\mathbf{k}}=(V_{1{\bf k}},\dots,V_{N{\bf k}})^T$ is the hole component of the eigenvector.
Here, we omit the band index of the BdG eigenvector, focusing on one particular BdG band. 


For each BdG band, we can construct the cell-periodic part of the quasiparticle wave function by combining the eigenvector of the BdG matrix and the cell-periodic part of the electron Bloch functions as
\begin{equation}\label{eq:eigenstate}
\psi_{  \mathbf{k}} ({\bf r})= \left[U_{{\bf k}}(\mathbf{r}) ,V_{\mathbf{k}}(\mathbf{r})\right]^T,
\end{equation}
where $U_{{\bf k}}(\mathbf{r})=\sum_{n}U_{n{\bf k}}[\phi_{n\mathbf{k}+\frac{\bf q}{2},\uparrow}({\bf r}),\phi_{n\mathbf{k}+\frac{\bf q}{2},\downarrow}({\bf r})]^T$ and $V_{\mathbf{k}}(\mathbf{r})=\sum_{n}V_{n{\bf k}}[\phi_{n-\mathbf{k}+\frac{\bf q}{2},\uparrow}^{*}({\bf r}),\phi^*_{n-\mathbf{k}+\frac{\bf q}{2},\downarrow}({\bf r})]^T$. 
We note that this wave function seems to show no difference from that of a homogeneous system; however, it contains the inhomogeneity of the perturbation fields through its dependency of $U$ and $V$ on the parameter ${\bf r}_c$. For future convenience, we define $\tilde\psi_{  \mathbf{k}} ({\bf r})= \left[U_{{\bf k}}(\mathbf{r}) e^{-i{\bf A}({\bf r}_c)\cdot {\bf r}} ,V_{\mathbf{k}}(\mathbf{r})e^{i{\bf A}({\bf r}_c)\cdot {\bf r}}\right]^T$ by putting the vector potential back into the cell-periodic part of the quasiparticle wave function.

\subsection{Quasiparticle wavepacket}\label{sub:wavepacket}
We can construct a wavepacket for superconducting quasiparticles with the quasiparticle wavefunctions of the local Hamiltonian \eqref{eq:eigenstate} \cite{sundaram1999wavepacket,liang2017wavepacket}
\begin{equation}
\Psi_{\mathbf{k}_{c}}({\bf r})=\int_{\bf k} a_{\mathbf{k},t} \tilde \psi_{{\bf k}}({\bf r}) e^{i{{\bf k}\cdot {\bf r}}},
\label{eq:wave packet}
\end{equation}
where $a_{\mathbf{k},t}$ is the constructing function, and $\int_{\bf k}$ is a shorthand of $\int d{\bf k}/(2\pi)^d$ with $d$ being space dimensionality, representing the momentum integral over the Brillouin zone. The constructing function $a_{\mathbf{k},t}$ is subject to several constraints. Firstly, the normalization of the wavepacket imposes a normalization for the constructing function as $\int_{\bf k}|  a_{{\bf k},t}|^{2}=1$. Secondly, the wavepacket is assumed to be sharply peaked at its center in momentum space $\mathbf{k}_c=\int_{\bf k}| a_{{\bf k},t}|^2 {\bf k}$ , which is the semiclassical momentum of the wavepacket. Therefore, the integration over the amplitude square of the constructing function combined with any other function will give its value at $\mathbf{k}_c$.
Thirdly, the wavepacket averaging over the position operator should be coincident with the center of the wavepacket, which requires that ${\bf r}_c=\langle \Psi_{\mathbf{k}_{c}}|\hat{\bf r}|\Psi_{\mathbf{k}_{c}}\rangle$.  
Substituting Eq.~(\ref{eq:wave packet}) into this quantum averaging, we can obtain the expression for the center of the wavepacket as
\begin{align}\label{eq:position}
    {\bf r}_c=&\langle\Psi_{{\bf k}_c}|\hat{\bf r}|\Psi_{{\bf k}_c}\rangle  \nonumber\\
    =&\int_{{\bf k},{\bf k'}} a_{\bf k}^*a_{\bf k'}\langle\tilde\psi_{\bf k}|e^{-i{\bf k}\cdot {\bf r}}(-i\nabla_{\bf k'}e^{i{\bf k}'\cdot {\bf r}})|\tilde\psi_{\bf k'}\rangle \nonumber\\
=&\nabla_{{\bf k}_c}\gamma_{{\bf k}_c}+\mathcal{A}_{{\bf k}_c},
\end{align}
where $\gamma_{{\bf k}}=-\arg(a_{{\bf k},t})$ is the phase of the constructing function and
\begin{equation}\label{eq:k-Berry}
\mathcal{A}_{\mathbf{k}} = \langle \psi_{\bf k}|i\partial_{\bf k} \psi_{\bf k} \rangle= \int_{\bf r} [U^\dagger_{{\bf k}}(\mathbf{r})  i\nabla_{\mathbf{k}} U_{{\bf k}}(\mathbf{r}) +V^\dagger_{{\bf k}}(\mathbf{r})  i\nabla_{\mathbf{k}} V_{{\bf k}}(\mathbf{r})]
\end{equation}
is the momentum-space superconducting Berry connection.
It is obvious that this Berry connection includes contributions from both the electron Bloch functions and the proximity-induced superconducting gap function.

We can obtain various physical quantities of the wavepacket through the quantum averaging of the relevant operator over the wavepacket (\ref{eq:wave packet}). We illustrate several examples here for future usage. Firstly, the effective charge of the wavepacket is given by
\begin{align}\label{eq:effective charge}
    \rho &=\langle\Psi_{\mathbf{k}_{c}}|{ \tau}_z|\Psi_{\mathbf{k}_{c}}\rangle =\langle\psi_{\mathbf{k}_{c}}|{ \tau}_z|\psi_{\mathbf{k}_{c}}\rangle
    \nonumber \\
    &= \int_{\bf r} [U^\dagger_{{\bf k}_c}(\mathbf{r})  U_{{\bf k}_c}(\mathbf{r}) - V^\dagger_{{\bf k}_c}(\mathbf{r})   V_{{\bf k}_c}(\mathbf{r}) ],
\end{align}
with $\tau_z$ being the Pauli matrix acting on the particle-hole space. Secondly, the $x$-component effective spin of the wavepacket is given by the quantum averaging,
\begin{align}
    {s}_x &=\langle\Psi_{\mathbf{k}_{c}}|{ \tau}_z \sigma_x|\Psi_{\mathbf{k}_{c}}\rangle =\langle\psi_{\mathbf{k}_{c}}|{ \tau}_z\sigma_x|\psi_{\mathbf{k}_{c}}\rangle
   \nonumber \\
    &= \int_{\bf r} [U^\dagger_{{\bf k}_c}(\mathbf{r})  \sigma_x U_{{\bf k}_c}(\mathbf{r}) - V^\dagger_{{\bf k}_c}(\mathbf{r})   \sigma_x  V_{{\bf k}_c}(\mathbf{r}) ],
\end{align}
where $\sigma_x$ is the Pauli matrix that is acting on the spin space. The $y$-component and $z$-component of the effective spin for the wavepacket are given similarly as ${s}_y =\langle \psi_{\mathbf{k}_{c}}|{\tau}_0\sigma_y| \psi_{\mathbf{k}_{c}}\rangle$ and ${s}_z =\langle\psi_{\mathbf{k}_{c}}|{\tau}_z\sigma_z| \psi_{\mathbf{k}_{c}}\rangle$. 

Thirdly, the charge and spin distributions on the wavepacket also have higher-order moments, such as the dipole moment \cite{xiao2010berry}. The charge dipole is written as,
\begin{equation}
    {\bf d}^{e}= \langle\Psi_{{\bf k}_c}|{\tau}_z(\hat{\bf r}-{\bf r}_c)|\Psi_{{\bf k}_c}\rangle= \mathcal{A}^{e}_{{\bf k}_c}-\rho\mathcal{A}_{{\bf k}_c},
\end{equation}
where we define the charge Berry connection as $ \mathcal{A}^e_{{\bf k}}= \langle \psi_{\bf k}|\tau_z|i\partial_{\bf k} \psi_{\bf k} \rangle=\int_{\bf r} [U^\dagger_{{\bf k}}(\mathbf{r})  i\nabla_{\mathbf{k}} U_{{\bf k}}(\mathbf{r}) -V^\dagger_{{\bf k}}(\mathbf{r})  i\nabla_{\mathbf{k}} V_{{\bf k}}(\mathbf{r})]$. This charge dipole is unique to superconducting systems because the effective charge is not a conserved quantity for superconducting quasiparticles. The spin dipole moment is calculated similarly. For example, the spin-$x$ dipole moment is written as,
\begin{equation}
    {\bf d}^{s_x}=\langle\Psi_{{\bf k}_c}|{\tau_z}{\sigma}_{x}(\hat{\bf r}-{\bf r}_c)|\Psi_{{\bf k}_c}\rangle= \mathcal{A}^{s_{x}}_{{\bf k}_c}-s_{x}\mathcal{A}_{{\bf k}_{c}},
\end{equation}
where we define the spin Berry connection for the x-component of the spin as $\mathcal{A}^{s_{x}} = \langle \psi_{\bf k}|\tau_z\sigma_x|i\partial_{\bf k} \psi_{\bf k} \rangle=\int_{\bf r} [U^\dagger_{{\bf k}}(\mathbf{r}) \sigma_x  i\nabla_{\mathbf{k}} U_{{\bf k}}(\mathbf{r}) -V^\dagger_{{\bf k}}(\mathbf{r}) \sigma_x i\nabla_{\mathbf{k}} V_{{\bf k}}(\mathbf{r})]$. The dipole moments for other spin components can be written with a similar procedure.


\subsection{Equations of motion}
Now we study the semiclassical dynamics of the wavepacket, which can be described by the equations of motion for the center of the wavepacket in both real and momentum space. The evolution of the wavepacket follows the Schr\"{o}dinger equation, which can be derived from the Lagrangian $\mathcal{L}[\Psi]=\langle\Psi|i\frac{d}{dt}-\hat{H}_c|\Psi\rangle$. The semiclassical Lagrangian is obtained by plugging the form of a wavepacket into this quantum Lagrangian
\begin{equation}
\mathcal{L}[{\bf r}_{c},{\bf k}_{c}]=\langle\Psi_{{\bf k}_c}|i\frac{d}{dt}-\hat{H}_c|\Psi_{{\bf k}_c}\rangle.
\end{equation}
According to Eqs. (\ref{eq:wave packet}) and (\ref{eq:position}), one has the wavepacket averaging over the time derivative as (See Appendix~\ref{app:lagrangian} for details):
\begin{align}\label{eq:time evolution}
   & \langle\Psi_{{\bf k}_c}|i\frac{d}{dt}|\Psi_{{\bf k}_c}\rangle=i\int _{\bf k}a_{\bf k}^{*}\dot{a}_{\bf k}
    +i\int_{{\bf k}} |a_{\bf k}|^2\langle\tilde\psi_{{\bf k}}\mid \dot{\tilde\psi}_{{\bf k}}\rangle
   \nonumber  \\
   &=\dot{\gamma}_{{\bf k}_c}-({\bf r}_c-\mathcal{A}_{{\bf k}_c})\cdot\dot{\bf k}_c+\mathcal{A}_{{\bf r}_c}\cdot\dot{\bf r}_c+\mathcal{A}_{t},
\end{align}
where $\mathcal{A}_{{\bf k}_c}$ is the momentum-space Berry connection (\ref{eq:momentum space Berry connection}) at the momentum ${\bf k}_c$ and we find the real-space Berry connection as
\begin{equation}
        \mathcal{A}_{{\bf r}_c}=\langle\psi_{{\bf k}_c}|i\nabla_{{\bf r}_c}\psi_{{\bf k}_c}\rangle+\frac{1}{2}(-\rho{\bf q}+{\bf B}\times{\bf d}^e),
\end{equation}
with ${\bf B}$ being the magnetic field and the time-space Berry connection as
\begin{equation}
    \mathcal{A}_{t}=\langle\psi_{{\bf k}_c}|i\partial_t\psi_{{\bf k}_c}\rangle.
\end{equation}
In deriving these Berry connections, we take a circular gauge and expand the gauge fields up to the first order.

The quantum averaging over the Hamiltonian is the eigen-energy of the local Hamiltonian, with the momentum at the center of the wavepacket
\begin{equation}\label{eq:energy}
   \langle\psi_{{\bf k}_c}|\hat{H}_c|\psi_{{\bf k}_c}\rangle = E_{{\bf k}_c}.
\end{equation}
Here we note that by taking the local Hamiltonian in calculating the quantum averaging, we have neglected the higher-order corrections, such as the contributions from the orbital magnetic moment.
 
Combining these results, we can write down the semiclassical action as $S[{\bf r}_c(t),{\bf k}_c(t)]=\int dt\mathcal{L}$ with the semiclassical Lagrangian
\begin{equation}\label{eq:Lagrangian}
\mathcal{L}= - E-\dot{\bf k}_c\cdot{\mathbf{r}}_c+\mathcal{A}_{{\bf k}_c}\cdot\dot{\bf k}_c+\mathcal{A}_{{\bf r}_c}\cdot\dot{\bf r}_c+\mathcal{A}_{t}, 
\end{equation} 
where the total time-derivative terms have been neglected. 
The equations of motion for the wavepacket are the Euler-Lagrangian equations for the semiclassical Lagrangian, which are written as
\begin{eqnarray}\label{eq:EOM}
        &\dot{{\bf r}}&=\nabla_{{\bf k}} E- \Omega_{{\bf k}{\bf k}}\cdot\dot{\bf k}-\Omega_{{\bf k}{\bf r}}\cdot\dot{\bf r}-\Omega_{{\bf k}t},\nonumber\\
&\dot{{\bf k}}&=-\nabla_{{\bf r}} E+ \Omega_{{\bf r}{\bf r}}\cdot\dot{\bf r}+ \Omega_{{\bf r}{\bf k}}\cdot\dot{\bf k}+\Omega_{{\bf r}t},
\end{eqnarray}
where we find various Berry curvatures given by the derivatives of the Berry connection as
\begin{equation}\label{eq:Berry}
    \Omega_{\alpha \beta}=\partial_{\alpha} \mathcal{A}_{\beta}-\partial_{\beta}\mathcal{A}_{\alpha}.
\end{equation}
Here, $\alpha,\beta = \{{\bf r}_c,{\bf k}_c, t\}$ denote all the possible components of real space, momentum space, and time. This equation formally resembles the standard equation of motion for band electrons given in Ref.~\cite{sundaram1999wavepacket}. However, the Berry curvatures have different physical pictures. The momentum space Berry curvature $\Omega_{\bf kk}$ is particularly interesting. It is an antisymmetric tensor and can be written in the form of a vector as $ \Omega^\alpha_k = \epsilon_{\alpha\beta\gamma}\Omega_{k_\beta k_\gamma}$. This momentum space Berry curvature vector can be explicitly written as
\begin{equation}
    {\bf\Omega}_{\bf k}=i\int_{\bf r}[\nabla_{\bf k}U^{\dagger}({\bf r})\times \nabla_{\bf k}U({\bf r})+\nabla_{\bf k}V^{\dagger}({\bf r})\times \nabla_{\bf k}V({\bf r})],
\end{equation}
where the cross product comes from the anti-symmetric nature of the tensor. The real-space Berry curvature can also be written as a vector similarly, while all other mixed Berry curvatures can only be expressed as antisymmetric tensors given in Eq. (\ref{eq:Berry}), which are demonstrated explicitly in Appendix~\ref{app:BC}. We will discuss the properties of two of the Berry curvatures in detail, which are useful for studying thermo-spintronics responses.

\subsection{Thermal responses}

The semiclassical approach is useful for studying various responses to dc and low-frequency perturbations \cite{xiao2010berry}. The thermal responses are of particular significance in the context of superconductors because the dc electric field is forbidden by the Meissner effect. Here, we apply the field variational approach established in Ref. \cite{dong2020berry} to formulate the thermal response for a general physical observable, where the mixed Berry curvatures defined in the pertinent parameter spaces are unveiled to play a vital role. 

In the wavepacket formalism, the local density of any physical observable ${\bf \theta}$ is given by
\begin{equation}\label{eq:density}
   {\bf \theta}({\bf r})=\int d{\bf r}_c\int_{{\bf k}_c}\mathcal{D}f\langle \Psi_{{\bf k}_c}|\hat{\bf \theta}\delta(\hat{\bf r}-{\bf r})|\Psi_{{\bf k}_c}\rangle,
\end{equation}
where $\mathcal{D}({\bf k}_c,{\bf r}_c)=1+\textrm{Tr}(\Omega_{{\bf k}_c{\bf r}_c})$ is the semiclassical phase space measure \cite{xiao2005berry}, and $f({\bf k}_c,{\bf r}_c)$ is the local equilibrium Fermi distribution function. Here we assume a scalar observable for simplicity, but it is straightforward to generalize the scheme for a vector or tensor observable. 
To calculate the local expectation with the field variational approach, we introduce a slowly varying auxiliary field $w^{\theta}({\bf r},t)$ that couples with the observable. This auxiliary field will be set to zero after we finish the calculation. The auxiliary Hamiltonian is written as
\begin{equation}
    \hat{\mathcal{H}}=\hat{H}_c+\hat{\bf \theta}w^{\theta}.
\end{equation}
The semiclassical action thus becomes a functional of the auxiliary field:
\begin{eqnarray}
    S[w^{\theta}({\bf r}_c,t)] = \int dt   \mathcal{L}, 
\end{eqnarray}
with $\mathcal{L}$ being the semiclassical Lagrangian, which takes the same form as Eq.~\eqref{eq:Lagrangian}. It was demonstrated that the local expectation of the operator can be expressed as the on-shell variation of the action with respect to the auxiliary field
\cite{dong2020berry}:
\begin{equation}\label{eq:variation}
   \langle\Psi_{{\bf k}_c}|\hat{\bf \theta}\delta(\hat{\bf r}-{\bf r})|\Psi_{{\bf k}_c}\rangle= -\frac{\delta S}{\delta w^{\theta}} |_{\text{onshell}},
\end{equation}
where the on-shell means that the wavepacket satisfies the Schr\"{o}dinger equation. We implement the resulting variation into the Eq. \eqref{eq:density} and obtain the local density of the observable
\begin{equation}
    {\bf \theta}({\bf r})=\int_{{\bf k}} \mathcal{D}f [\partial_{ w^{\theta}}\Tilde{E}-{\Omega}_{w^{\theta}{ k}_{\alpha}}\cdot\dot{k}_{\alpha}-{\Omega}_{w^{\theta}{ r_{\alpha}}}\cdot\dot{{r}}_{\alpha}]-\partial_{r_{\alpha}}\int_{{\bf k}}f{d}^{\theta}_{\alpha}.
\end{equation}
Here, $\Omega_{w^{\theta}{ k}_{\alpha}}$ and $\Omega_{w^{\theta}{r_{\alpha}}}$ are mixed Berry curvatures in the space expanded by $\{{\bf r},{\bf k},{ w^{\theta}}\}$, taking the same form as the Eq.~\eqref{eq:Berry}. $\Tilde{E}=E+\Delta E$ is the particle energy, with $\Delta E={d}^{\theta}_{\alpha}\partial_{r_{\alpha}}{ w^{\theta}}$ being the gradient correction arising from the $\theta$ dipole moment 
\begin{equation}
{ d}^{ \theta}_{\alpha}=\langle\Psi_{{\bf k}_c}|\hat{\theta}(\hat{\bf r}-{\bf r}_c)_{\alpha}|\Psi_{{\bf k}_c}\rangle.
\end{equation}
Here and hereafter, $\alpha$ and $\beta$ denote the spatial directions, and we take the Einstein convention in summing repeated labels. 

Now we can plug the semiclassical equations of motion \eqref{eq:EOM} into the above formula. With some tedious manipulation of the expressions and taking advantage of the Bianchi identity $\partial _{w^{\theta}}\Omega_{k_{\alpha}r_{\alpha}}+\partial_{r_{\alpha}}\Omega_{w^{\theta}k_{\alpha}}+\partial_{k_{\alpha}}\Omega_{r_{\alpha}w^{\theta}}=0$, the local density can be evaluated up to the first order of the spatial gradient. The result can be separated into the equilibrium and off-equilibrium parts:
\begin{equation}\label{eq:local density}
    \theta_{\text{eq}}=  \theta_0-\partial_{r_\alpha} { D}_{\alpha}^{\bf\theta},\quad
   \theta_{\text{neq}}= -{\bf \alpha}^{\theta}_{\alpha}\partial_{{ r_{\alpha}}}T.
\end{equation}
The equilibrium density is already present at equilibrium, which incorporates a monopole density $\theta_0$ as well as the divergence of a dipole density. The monopole density and dipole density are given by
\begin{equation}\label{eq:monopole}
    \theta_0=\partial_{w^\theta}\int_{{\bf k}}\mathcal{D}g(\Tilde{E}), \quad {D}_{\alpha}^{\bf\theta}=\int_{{\bf k}}(f{d}_{\alpha}^{\theta}+g\Omega_{{ k}_{\alpha}w^{\theta}}),
\end{equation}
where  $g=-k_{B}T\ln(1-f)$ is the grand canonical potential density for a state. The off-equilibrium part is linearly dependent on the temperature gradient, with the response coefficient given by 
\begin{equation}\label{eq:response}
    \alpha_{\alpha}^{\theta}=\int_{{\bf k}} \mathfrak{s}\Omega_{{ k}_{\alpha}w^\theta}|_{w^\theta\to 0},
\end{equation}
where $\mathfrak{s}=-\partial g/\partial T$ is the state-resolved entropy density. The response is dominated by the mixed Berry curvature $\Omega_{{ k}_{\alpha}w^\theta}$, which can be cast into (after taking $w^\theta\to 0$)
\begin{equation}
\Omega_{k_{\alpha}w^\theta}=-2\textrm{Im}\sum_{m\ne n}\frac{{v}_{\alpha,nm}{\theta}_{mn}}{(E_m-E_n)^2},
\end{equation}
where $v_{\alpha,nm}=\langle\psi_n|\hat{v}_{\alpha}|\psi_m\rangle$ is the inter-band matrix element of the velocity operator with the quasiparticle velocity operator $\hat{v}_{\alpha}=\partial H_{\bf k}/\partial k_{\alpha}$, and $\theta_{mn}=\langle\psi_m|\hat{\theta}|\psi_n\rangle$ is the inter-band matrix element of the operator $\hat\theta$ for the physical observable. This formula is gauge-invariant. Therefore, it is convenient for numerical calculations.

\section{Thermo-spintronic responses of Berry curvature origins}\label{sec:response}
Thermo-spintronic responses of Berry curvature origins in superconductors have not been studied before. In this work, we study two of such effects: the thermal Edelstein effect and the spin Nernst effect. 

\subsection{Thermal Edelstein effect}
The well-known Edelstein effect refers to a spintronic phenomenon in which the electric field induces a spin polarization \cite{edelstein1990spin}. The thermal counterpart of this phenomenon is the thermal Edelstein effect \cite{freimuth2014DMI-SOT,xiao2016spin,freimuth2016inverse,Shitade2019,dong2020berry}, which describes the spin polarization driven by a temperature gradient
\begin{equation}\label{eq:edelstin}
     \delta S_{\beta}=\chi_{\alpha\beta}(-\nabla_{{\alpha}} T),
\end{equation}
where $\delta s_{\beta}$ is the spin generation in $\beta$ direction,  $-\nabla_{{\alpha}} T$ is the temperature gradient along $\alpha$ direction, and $\chi_{\alpha\beta}$ is the thermal Edelstein coefficient.

The thermal Edelstein coefficient can be easily calculated using the semiclassical theory established in the last section. In Nambu representation, the spin operators are defined as $\hat{\mathbf s} = (\tau_z\sigma_x,\tau_0\sigma_y,\tau_z\sigma_z)$. We introduce an auxiliary Zeeman field ${\bf m}$, which couples with the spin operator, and introduce an auxiliary Zeeman energy to the Hamiltonian. Then the thermal Edelstein coefficient is given by Eq. \eqref{eq:response} as
\begin{equation}\label{thermoEdelstein}
\chi_{\alpha\beta}=\int_{\bf k}\mathfrak{s}\Omega_{k_{\alpha}m_{\beta}}|_{{\bf m}\to 0}.
\end{equation}
Here $\Omega_{k_{\alpha}m_{\beta}}$ is the $\alpha\beta$ component of the mixed Berry curvature defined in the phase space expanded by momentum ${\bf k}$ and auxiliary Zeeman field ${\bf m}$.

\subsection{Spin Nernst effect}
The spin Nernst effect describes the transverse spin current driven by a longitudinal temperature gradient \cite{Bose2018SNE,Bose2018SNT,kim2020SNT,zhang2020SNE,jain2023thermally}. 
Its derivation is more involved due to the nonconserved nature of the spin in SOC systems. Here, we adopt the definition of the bulk conserved spin current \cite{shi2006proper,Murakami2006conserved,Xiao2018SNE,xiao2021conserved}, 
which incorporates not only the conventional spin current but also an additional contribution from the spin torque.

Conventionally, the spin current operator is defined as the symmetrized product of the spin operator and  the velocity operator $\hat{ J}^{s}=\frac{1}{2}\{\hat{ s},\hat{ v}\}$. Under this definition, there will be a source term in the continuity equation for spin (assume steady state thus $\partial_t{s}=0$)
\begin{equation}\label{eq:continuity}
    \nabla\cdot {\bf J}^{ s}_{}={\tau},
\end{equation}
where ${\bf J}^{s}$ represents the conventional spin current density.
The source term ${\tau}$, characterizing the spin precession during the propagation, is the spin torque density, whose operator is given by the time derivative of the spin operator $\hat{\tau}=-i[\hat{ s},\hat{H}]$. 

For the conventional spin current, its local density can be evaluated up to the first order of spatial gradient as Eq. \eqref{eq:local density}:
\begin{eqnarray}
    { J}^{s}_{\text{eq},\alpha}={ J}^{ s}_{0,\alpha}-\partial_{r_{\beta}} \cdot { D}^{{ J}^s_{\alpha}}_{\beta},\quad { J}^{s}_{\text{neq},\alpha}=-\alpha^{{ J_{\alpha}^s}}_{\beta}\cdot\partial_{r_{\beta}} T.
\end{eqnarray}
The attained thermal response coefficient is given by 
\begin{equation}
    \alpha^{{ J_{\alpha}^s}}_{\beta}=\int_{{\bf k}}\mathfrak{s} \Omega_{k_{\beta}w_{\alpha}^{J^{s}}}|_{w_{\alpha}^{J^{s}}\to 0},
\end{equation}
where the corresponding mixed Berry curvature is explicitly written as 
\begin{equation}
  \Omega_{k_{\beta}w^{J^s}_{\alpha}}=-2\textrm{Im}\sum_{m\ne n}\frac{{v}_{\beta,nm}{J}_{\alpha,mn}}{(E_m-E_n)^2},
\end{equation}
with $w_{\alpha}^{J^{s}}$ denoting the auxiliary field that couples with the conventional spin current $J^s_{\alpha}$ \cite{dong2020berry}.  
The equilibrium conventional spin current density is determined by its monopole density and dipole density, which are evaluated respectively as 
\begin{equation}\label{eq:conventional spin current at eq}
    {J}^{ s}_{0,\alpha}=\int_{\bf k}f\langle\psi_{\bf k}|\hat{J}^{s}_{\alpha}|\psi_{\bf k}\rangle,
    \quad { D}^{{ J}_{\alpha}^s}_{\beta}=\int _{{\bf k}}(f{ d}^{{ J}_{\alpha}^s}_{\beta}+g\Omega_{k_{\beta}w_{\alpha}^{J^{s}}}),
\end{equation}
where ${ d}_{\beta}^{ J_{\alpha}^s}$ represents the spin current dipole moment. 
In the presence of SOC, however, the equilibrium current density ${ J}_{0,\alpha}^{s}$ can be nonzero in a uniform system, and the dipole density induced current may not be circulating, which can not serve as a magnetization current. So the attained conventional spin current density may have a non-vanishing net flow at equilibrium \cite{xiao2021conserved}.

For the torque density, its monopole part in the multipole expansion must vanish at steady state, because it is a total time derivative of a local operator (i.e., spin) \cite{Sugimoto2006PRB,xiao2021conserved}. Therefore, one can always expand the torque density as 
\begin{equation}
    \tau=-\nabla\cdot{\bf J}^{\tau}.
    \label{torque current}
\end{equation}
To evaluate ${\bf J}^{\tau}$ in nonuiform cases, second-order expansion for the torque density is inevitable. This theory is parallel to that developed in Ref. \cite{xiao2021conserved}. Therefore, here we focus on the physical consequences.
The torque-induced spin current can be separated into the equilibrium and nonequilibrium parts as
\begin{equation}
    {J}^{\tau}_{\text{eq},\alpha}=D_{\alpha}^{\tau}-\partial_{r_{\beta}}Q^{\tau}_{\alpha\beta},\quad {J}^{\tau}_{\text{neq},\alpha}=-\alpha^{{\bf J}^{\tau}}_{\alpha\beta}\partial_{r_{\beta}}T.
\end{equation}
The torque-induced equilibrium spin current stems from the torque dipole density and the torque quadrupole density. They are given by
\begin{equation}
    D^{\tau}_{\alpha}=\int_{\bf k}(fd^{\tau}_{\alpha}+g\Omega_{k_{\alpha}w^{\tau}}),\quad   Q^{\tau}_{\alpha\beta}=\int_{{\bf k}}(fq_{\alpha\beta}^{\tau}+g\chi_{\alpha\beta}^{\tau}).
\end{equation}
The torque dipole density is determined by the dipole moment $d^{\tau}_{\alpha}$ of each wavepacket and the mixed Berry curvature
\begin{equation}
\Omega_{k_{\alpha}w^{\tau}}=-2\textrm{Im}\sum_{m\ne n}\frac{{v}_{\alpha,nm}{\tau}_{mn}}{(E_m-E_n)^2},
\end{equation}
with $w^{\tau}$ as the auxiliary field that couples to the torque operator. 
The torque quadrupole density is determined by the quadrupole moment of each wavepacket $  {q}^{\tau}_{\alpha\beta}=\langle\Psi|\hat{\tau}(\hat{\bf r}-{\bf r}_c)_{\alpha}(\hat{\bf r}-{\bf r}_c)_{\beta}|\Psi\rangle$ and
\begin{equation}
    \chi_{\alpha\beta}^{\tau}=-2\textrm{Im}\sum_{m\ne n}\frac{v_{\beta,nm}d^{\tau}_{\alpha,mn}}{(E_m-E_n)^2} +\partial_{k_{\alpha}}\mathcal{G}_{k_{\beta}w^{\tau}},
\end{equation}
with
\begin{equation}
    \mathcal{G}_{k_{\beta}w^{\tau}}=\textrm{Re}\sum_{m\ne n}\frac{{v}_{\alpha,nm}{\tau}_{mn}}{(E_m-E_n)^2} 
\end{equation}
being a mixed quantum geometric metric.
As for the transport nonequilibrium spin current, the corresponding spin Nernst conductivity is given by
\begin{equation}
    \alpha^{{\bf J}^{\tau}}_{\alpha\beta}=\int_{{\bf k}}\mathfrak{s}\chi^{\tau}_{\alpha\beta}.
\end{equation}

According to Eqs. (\ref{eq:continuity}) and (\ref{torque current}), the conserved spin current is given by
\begin{equation}
\boldsymbol{\mathcal{J}}^{ s}={\bf J}^{\ s}+{\bf J}^{\tau}.
\end{equation}
The equilibrium part of the conserved spin current reads
\begin{equation}
    \mathcal{J}^{s}_{\text{eq},\alpha}={J}^{s}_{0,\alpha}+D_{\alpha}^{\tau}-\partial_{r_{\beta}}(D_{\beta}^{J_{\alpha}}+Q_{\alpha\beta}^{\tau}).
\end{equation}
The transport spin current driven by the thermal gradient is 
\begin{equation}
    \mathcal{J}^{s}_{\text{neq},\alpha}=-\alpha_{\alpha\beta}\partial_{\beta}T,\quad \alpha_{\alpha\beta}=\int_{{\bf k}}\mathfrak{s}(\Omega_{k_{\beta}w_{\alpha}^{J^{s}}}+\chi_{\alpha\beta}^{\tau}).
\end{equation}
By examining the relevant quantities in detail, one finds \cite{Xiao2021conserved-arxiv}
\begin{eqnarray}
   && d^{\tau}_{\alpha}=-J^{s}_{\alpha}+v_{\alpha}s,\quad  \Omega_{k_{\alpha}w^{\tau}}=\partial_{k_{\alpha}}s,\quad\mathcal{G}_{k_{\alpha}w^{\tau}}=d^{s}_{\alpha},\nonumber\\   &&\chi^{\tau}_{\alpha\beta}=-\Omega_{k_{\beta}w_{\alpha}^{J^{s}}}+\Omega_{k_{\beta}k_{\alpha}}s+\partial_{k_{\beta}}d^{s}_{\alpha},\nonumber\\ 
  &&  q^{\tau}_{\alpha\beta}=-{ d}_{\beta}^{ J_{\alpha}^s}+d^{s}_{\alpha}v_{\beta}+m^{s}_{\beta\alpha},
\end{eqnarray}
where $m^{s}_{\beta\alpha}=\epsilon_{\beta\alpha\gamma}m^s_\gamma$ is an antisymmetric tensor dual to the vector 
\begin{equation}
    {\bf m}^s=\frac{1}{2}\textrm{Re}\sum_{m\ne n}\mathcal{A}_{nm}\times J^s_{mn}+\frac{1}{2}{\bf d}^s\times {\bf v},
\end{equation}
with $\mathcal{A}_{nm}=\langle\psi_{n{\bf k}}|i\nabla_{\bf k}\psi_{m{\bf k}}\rangle$ being the interband Berry connection in momentum space and ${\bf v}$ being the quasiparticle velocity expectation.
Employing these relations, we find
\begin{eqnarray}
\boldsymbol{\mathcal{J}}^{s}_{\text{eq}}&=&\nabla\times{\bf M}^{s}
\end{eqnarray}
is indeed a circulating magnetization current, with the vector
\begin{eqnarray}
{\bf M}^{s}=\int_{\bf k} (f{\bf m}^s+g\Omega_{\bf k}s)
\end{eqnarray}
interpreted as the orbital magnetization of spin. Subtracting this circulating current from the total spin current leads to the transport spin current
\begin{eqnarray}\label{eq:intrinsic nernst}
        \alpha_{\alpha\beta}&=&\int_{{\bf k}}\mathfrak{s}(\Omega_{k_{\beta}k_{\alpha}}s+\partial_{k_{\beta}}d_{\alpha}^{s}).
\end{eqnarray}

Moreover, as shown in Ref. \cite{xiao2021conserved}, the second term in the above equation is eventually canceled out by the accumulation of spin-dipole variation during scattering of BdG particles with impurities. The latter gives rise to a spin current 
\begin{equation}\label{eq:scatter}
    \boldsymbol{\mathcal{J}}^{s}_{\text{dp}}=\int_{\bf k}\delta f_{n\bf k}\int_{\bf k'}P^{n'n}_{\bf k' k}({\bf d}^s_{n'{\bf k'}}-{\bf d}^s_{n{\bf k}}),
\end{equation}
where $\delta f_{n\bf k}$ is the off-equilibrium distribution determined by the Boltzmann transport equation
\begin{equation}
    \frac{E_k}{T}\partial_{r_{\alpha}}T\partial_{k_\alpha}f=\int_{\bf k'}P^{n'n}_{\bf k' k}(\delta f_{n'\bf k'}-\delta f_{n\bf k}),
    \label{SBE}
\end{equation}
and $P^{n'n}_{\bf k'k}$ is the scattering rate. Substituting Eq.~(\ref{SBE}) into (\ref{eq:scatter}) and making use of $\partial_{k_{\beta}}\mathfrak{s}=\frac{E_k}{T}\partial_{k_\alpha}f$, we obtain 
\begin{equation}
\boldsymbol{\mathcal{J}}^{s}_{\text{dp}}=-\partial_{r_{\beta}}T\int_{\bf k} \mathfrak{s} \partial_{k_{\beta}}{\bf d}^s_{},
\end{equation}
which turns out to be completely independent of the scattering. This contribution cancels the second term in Eq. \eqref{eq:intrinsic nernst}, yielding a pure Hall-type spin current response
\begin{equation}\label{eq:Nernst}
\boldsymbol{\mathcal{J}}_{H}^{s_{\alpha}}=\alpha_{H}^{\alpha}\times(-\nabla T),
\end{equation}
where$ \boldsymbol{\mathcal{J}}_{H}^{s_{\alpha}}$ is the transverse spin current of the $\alpha$-component spin, and 
\begin{equation}
\mathbf{\alpha}^{\alpha}_H = \int_{\mathbf{k}} \mathfrak{s} \Omega_{\mathbf{k}}s_{\alpha}
\label{eq:ASN}
\end{equation}
denotes the corresponding spin Nernst conductivity.

\section{Analysis of Berry curvature }\label{sec:curvature}
The Berry curvature around the electron Fermi surface is particularly interesting because it is most relevant to nonequilibrium quasiparticle response as long as the superconducting gap is the smallest energy scale of the system, which is usually indeed the case for realistic proximity-induced superconducting systems. Fortunately, this Berry curvature can be calculated analytically because the possibly complicated BdG bands are simplified around the electron Fermi surface. In the following, we present this analysis with both general formulas and results of specific models. 

Around each electron Fermi surface, the complicated BdG matrix can be reduced to a two-by-two matrix, where the diagonal elements are the electron and hole dispersions of the relevant band, and the off-diagonal element is the intra-band component of the effective gap in the band representation (\ref{eq:gapfunction}).
Then the BdG equation (\ref{eq:BdG}) reduces to
\begin{equation}
\left(\begin{array}{cc}
\xi_{n}\left(\mathbf{k}\right) & \tilde{\Delta}_{nn}\left(\mathbf{k}\right)\\
\tilde{\Delta}_{nn}^{*}\left(\mathbf{k}\right) & -\xi_n\left(-\mathbf{k}\right)
\end{array}\right)\left(\begin{array}{c}
u\\
v
\end{array}\right)=E\left(\begin{array}{c}
u\\
v
\end{array}\right),
\end{equation}
where $u$ and $v$ are the eigenfunctions, $E$ is the eigenvalue, $\xi_{n}({\bf k})$ is the electron dispersion around the Fermi surface, and the intra-band effective gap is written as
\begin{equation}\label{eq:intra-band gap}
     \tilde{\Delta}_{nn{\bf k}}
= \Delta_{{\bf k}}\int_{{\bf R},{{\bf r}}} \phi_{n\mathbf{k},\sigma}^{*}({\bf R}+\frac{\bf r}{2}) \chi^c_{\sigma\sigma'}({\bf r})
\phi_{n-\mathbf{k},\sigma'}^{*}({\bf R}-\frac{\bf r}{2}). 
\end{equation}
This reduced BdG equation can be analytically solved with the eigenfunctions
\begin{equation}
    u=e^{\frac{i}{2}\varphi}\sqrt{\frac{1}{2}(1+\rho)},\quad v=e^{-\frac{i}{2}\varphi}\sqrt{\frac{1}{2}(1-\rho)},
\end{equation}
where $\varphi = \arg(\tilde{\Delta}_{nn})$ is the phase of the effective gap, and $\rho=\xi_{n}/\sqrt{\xi^2_{n} + |\tilde{\Delta}_{nn}|^2}$ is the effective charge of the superconducting quasiparticle. With these eigenfunctions, we can construct the approximate quasiparticle eigenstates as
\begin{equation}\label{eq:WFEFS}
|\psi_{n\mathbf{k}}\rangle=[u|\phi_{n\mathbf{k}}\rangle,v|\phi^*_{n-\mathbf{k}}\rangle]^T,
\end{equation}
where $|\phi_{n{\bf k}}\rangle$ represents the cell-periodic Bloch wave function of the electron band relevant to the Fermi surface.

\subsection{Berry curvature around the electron Fermi surface}
The approximated quasiparticle eigenstate around the electron Fermi surface enables the division of the Berry connections and Berry curvatures into the contributions from the effective superconducting gap function and the contributions from the geometric properties of the electron Bloch function.
We first examine the momentum space Berry connection, which is written as
\begin{eqnarray}\label{eq:momentum space Berry connection}
\mathcal{A}_{{\bf k}}  =& &\langle\phi_{n\mathbf{k}}| (u^{*}i\nabla_{\mathbf{k}}u)|\phi_{n\mathbf{k}}\rangle
+\langle\phi^*_{n-\mathbf{k}}|(v^{*}i\nabla_{\mathbf{k}}v)|\phi_{n-\mathbf{k}}^{*}\rangle
\nonumber\\
&&+|u|^{2}\langle\phi_{n\mathbf{k}}| i\nabla_{\mathbf{k}}\phi_{n\mathbf{k}}\rangle
+| v|^{2}\langle\phi^*_{n-\mathbf{k}}|i\nabla_{\mathbf{k}}\phi_{n-\mathbf{k}}^{*}\rangle
\nonumber\\
 =&&  -\frac{\rho}{2} \nabla_{\bf k}\varphi+| u|^{2}\mathcal{A}^{b}_{n}({\bf k})+| v|^{2}\mathcal{A}^{b}_{n}(-{\bf k}),
\end{eqnarray}
where $\mathcal{A}^{b}_{n}({\bf k})=\langle \phi_{n{\bf k}}|i\nabla_{\mathbf{k}}\phi_{n{\bf k}}\rangle$ is the electron Berry connection. Then the Berry curvature is expressed as the combination of two contributions,
\begin{equation}\label{eq:BCFS}
\mathbf{\Omega_{k}}=\frac{1}{2}  D_{\bf k}\varphi \times \nabla_{\mathbf{k}}\rho+| u|^{2}\mathbf{\Omega}^{b}_{n} ({\bf k})- |v|^{2}\mathbf{\Omega}^{b}_{n}(-{\bf k}),
\end{equation}
where $\mathbf{\Omega}_{n}^{b}({\bf k})=\nabla_{\bf k}\times \mathcal{A}_{n}^{b}({\bf k})$ in the second and third terms are exactly the electron Berry curvature. 

The first term in the Berry curvature~\eqref{eq:BCFS} is expressed as the cross product of $k$-derivative of the effective charge $\nabla_{\bf k}\rho$ and a gauge-invariant $k$-derivative of the superconducting phase
\begin{equation}\label{eq:Dk phi}
  D_{\bf k}\varphi=\nabla_{\bf k}\varphi-\mathcal{A}^{b}_{n}({\bf k})+\mathcal{A}^{b}_{n}(-{\bf k}).
\end{equation}
$D_{\bf k}\varphi$ formally resembles the supervelocity in real space $D\varphi=\nabla\varphi-2{\bf A}$. Its gauge invariance can be easily checked: any gauge transformation on the electron wave function $\phi_{\bf k}\to e^{i\lambda(\mathbf{k})} \phi_{\bf k}$ will simultaneously transform the electron band Berry connection as $\mathcal{A}^{b}_{n}({\bf k})\to\mathcal{A}^{b}_{n}({\bf k})-\nabla_{\mathbf{k}}\lambda(\mathbf{k})$ and the effective gap phase as $\varphi(\mathbf{k}) \to \varphi(\mathbf{k}) -[\lambda(\mathbf{k})+\lambda(\mathbf{-k})]$. As a result, the extra terms from gauge freedom cancel each other in the gauge invariant $k$-derivative of the phase $D_{\bf k}\varphi$.

The second and three terms in the Berry curvature~\eqref{eq:BCFS} also manifest the entanglement of the superconducting pairing and the electron band geometry. The co-presence of the electron geometric quantities at $\bf k$ and $-\bf k$ in the equation for the Berry connection and Berry curvature uncovers the basic property that the superconducting quasiparticle is a quantum superposition of the electron at ${\bf k}$ and the hole at $-{\bf k}$.

Next, we follow the previous treatment for the momentum space Berry curvature to derive a formula for the $\bf m$-space superconducting Berry connection around the electron Fermi surface, which is expressed as
\begin{equation}
    \mathfrak{A} ({\bf k})=-\frac{\rho}{2}\nabla_{\bf m}\varphi + |u|^2\mathfrak{A}_{n}^b(\mathbf{k})-|v|^2\mathfrak{A}_{n}^b(-\mathbf{k}),
\end{equation}
where $\mathfrak{A}_{n}^b(\mathbf{k})=\langle \phi_{n\mathbf{k}}|i\nabla_{\bf m}\phi_{n\mathbf{k}}\rangle$ is the electron Berry connection in the $\bf m$-space. Combining the momentum-space Berry connection and this $\bf m$-space Berry connection, the mixed phase space superconducting Berry curvature can be written as
\begin{align}\label{eq:phase-space BC}
    \Omega_{ k_{\alpha}m_{\beta}}
    &=\partial_{k_{\alpha}}\mathfrak{A}_{m_{\beta}}-\partial_{ m_{\beta}}\mathcal{A}_{k_{\alpha}}
   \nonumber \\
    &=\frac{1}{2}(D_{k_{\alpha}}\varphi \partial_{m_{\beta}}\rho-D_{m_{\beta}}\varphi \partial_{k_{\alpha}}\rho)\nonumber\\
    &+
|u|^2\Omega^{b}_{ k_{\alpha}m_{\beta}}(\mathbf{k})+|v|^2\Omega^{b}_{ k_{\alpha}m_{\beta}}(-\mathbf{k}),
\end{align}
where we find the $\bf m$-space counterpart of the gauge-invariant derivative of the superconducting phase defined as
\begin{equation}
    D_{\bf m}\varphi=\nabla_{\bf m}\varphi-\mathfrak{A}_{n}^{b}(\mathbf{k})-\mathfrak{A}_{n}^{b}(-\mathbf{k}).
\end{equation}
This expression demonstrates the universality of the gauge-invariant derivative of the superconducting phase in characterizing the Berry curvatures of superconducting systems.

\subsection{Two-band electron system}
The formulas for the Berry curvatures around the electron Fermi surface can be applied to the simplest model where the electron has a two-by-two Hamiltonian and the pairing function is even parity. 
For two-band electron systems, the Hamiltonian can be written without losing generality as
\begin{equation}\label{eq:electron H}
    \mathcal{H}_{0}=  h_0 +  \mathbf{h}\cdot\boldsymbol{\sigma},
\end{equation}
where $h_0=\epsilon_{\bf k} - \mu $ is the  electron kinetic energy subtracting the chemical potential $\mu$, and $\mathbf{h}=(h_x,h_y,h_z)$ is a three-dimensional vector coupling with the Pauli matrices vector $\boldsymbol{\sigma}=(\sigma_{x},\sigma_{y},\sigma_{z})$. We assume that this electron Hamiltonian has $C_{2z}$ symmetry, and the corresponding electron spectrum reads
\begin{equation}
    \xi_{\mathbf{k}}= h_0 \pm h,
\end{equation}
where the $\pm$ sign denotes the two electron bands, and the zeros of the electron spectrum $\xi({\mathbf{k}_F}) = 0$ give the electron Fermi surface ${\mathbf{k}_F}$.
The eigenfunction of the electron Hamiltonian can be analytically expressed as $|\phi_{+,\mathbf{k}}\rangle = (\cos\frac{\theta}{2}e^{-i\zeta},\sin\frac{\theta}{2})^T$ for the upper band and $|\phi_{-,\bf k}\rangle=(\sin \frac{\theta}{2} ,-\cos\frac{\theta}{2}e^{i\zeta})^T$ for the lower band, where $\theta=\arccos(h_z/h)$ and $\zeta=\arctan(h_y/h_x)$ are the polar and the azimuthal angles of the vector ${\bf h}$, respectively.
When the electron system is in proximity to a superconductor, this two-band electron system acquires a proximity-induced superconducting gap $\Delta_{\bf k}$. The BdG Hamiltonian is a four-by-four matrix which is written as
\begin{equation}\label{eq:2bandmodel}
  {H}_{\mathbf{k}}=\left(\begin{array}{cc}
\mathcal{H}_{0,{\bf k}} & i\Delta_{\bf k}\sigma_{y}\\
-i\Delta^*_{\bf k} \sigma_{y} & -\mathcal{H}_{0,\mathbf{-k}}^{*}
\end{array}\right),
\end{equation} 
where the proximity-induced gap $\Delta_{\bf k}$ can possess a non-vanishing phase $\eta_{\bf k}$ when the proximity-induced superconducting gap is a chiral gap function. For a simple $s$-wave or $d$-wave superconducting gap, we can simply set $\eta_{\bf k}$ as a constant all around the Brillouin zone.

The intra-band effective gap is shaped by the electron states of opposite momentum, as was shown in Eq. (\ref{eq:intra-band gap}). For this two-band electron system, the expression for the effective gap can be analytically expressed as
\begin{equation}\label{eq:2band intra-gap}
    \tilde{\Delta}_{\bf k}=\bar {\Delta}_{\bf k}\sin\theta e^{i(\eta_{\bf k}\pm \zeta)},
\end{equation}
where $\bar {\Delta}_{\bf k}$ is the amplitude part of ${\Delta}_{\bf k}$.
Compared with the proximity-induced gap, the effective gap acquires an amplitude modulation of $\sin\theta$ and an extra phase of $\zeta$ from the electron bands. Since the electron Berry connection of a two-band system is simply given as $\mathcal{A}^{b}_{\bf k}({\bf k})=\pm \cos^{2}\frac{\theta}{2}\nabla_{\bf k}\zeta$ and $\mathcal{A}^{b}_{\bf k}(-{\bf k})=\mp \cos^{2}\frac{\theta}{2}\nabla_{\bf k}\zeta$, we can plug the phase of the effective gap into Eq.~\eqref{eq:Dk phi} and derive the gauge invariant $k$-derivative of the superconducting phase as
\begin{equation}\label{eq:D k phi for two band model}
    D_{\bf k}\varphi=\nabla_{\bf k}\eta_{\bf k}\mp\cos\theta\nabla_{\bf k}\zeta.
\end{equation}
Now we can write down the superconducting Berry curvature given by Eq.~\eqref{eq:BCFS} as
\begin{eqnarray}\label{eq:Berry curvature for two band model}
{\bf \Omega}_{\bf k}
    =\frac{1}{2}(\nabla_{\mathbf{k}}\eta_{\bf k}   \mp \cos\theta\nabla_{\mathbf{k}}\zeta )\times \nabla_{\mathbf{k}}\rho  + \rho{\bf \Omega}_{\bf k}^{b},
\end{eqnarray}
where $\rho = \xi_{\bf k} / E_{\bf k}$ is the effective charge of the quasiparticle wavepacket with $E_{\bf k } = \sqrt{ \xi^2 _{\bf k} + {\tilde \Delta}^2_{\bf k}}$ being the quasiparticle spectrum around the electron Fermi surface, and $\Omega_{\bf k}^b=\pm {\mathbf{h}}\cdot(\partial_{k_x}\mathbf{h}\times\partial_{k_y}\mathbf{h})/{h^3}$ is the electron Berry curvature.

This expression for the superconducting Berry curvature is decomposed into three parts. The first part $\frac{1}{2}\nabla_{\mathbf{k}}\eta_{\bf k} \times \nabla_{\mathbf{k}}\rho$ denotes the Berry curvature purely stemming from the intrinsic chirality of the proximity-induced superconducting gap, such as the chiral $d$-wave pairing \cite{wang2021berry}. The second term $ \mp\frac{1}{2}  \cos\theta\nabla_{\mathbf{k}}\zeta \times \nabla_{\mathbf{k}}\rho $ is the Berry curvature stemming from the effective chiral pairing induced by the geometry of the electron band. This term resembles the one from the intrinsic chiral pairing gap, but with an extra modulating function of $\cos \theta$. The third term $\rho{\bf \Omega}_{\bf k}^{b}$ is simply the residue of the electron Berry curvature. This term becomes relatively large when the Fermi surface gets close to the top or bottom of the electron band.

For the two-band electron model, the analysis for the momentum-space Berry curvature can be easily extended to study the mixed phase space Berry curvature that dominates the thermal Edelstein effect. We consider the same two-band electron Hamiltonian as in Eq.~(\ref{eq:2bandmodel}), and introduce an auxiliary field $\mathbf{m}$ which couples with the spin, inducing an auxiliary Zeeman energy ${\bf m}\cdot \boldsymbol{\sigma}$. This auxiliary field modulates the BdG Hamiltonian
\begin{equation}
   {H}_{\mathbf{k}}=\left(\begin{array}{cc}
\mathcal{H}_{0,{\bf k}}+{\bf m}\cdot \boldsymbol{\sigma} & i\Delta_{\bf k}\sigma_{y}\\
-i\Delta^*_{\bf k} \sigma_{y} & -\mathcal{H}_{0,\mathbf{-k}}^{*}-{\bf m}\cdot  \boldsymbol{\sigma}^*
\end{array}\right).
\end{equation} 
We focus on the in-plane spin responses. While the in-plane components of $\mathbf{m}$ break the $C_{2z}$ symmetry of the electron Hamiltonian, the constraint on the relation between $\mathbf{h}({\bf k})$ and $\mathbf{h}(-{\bf k})$ still holds, which writes as $\mathbf{h}(-{\bf k})=(-h_x(\mathbf{k}),-h_y(\mathbf{k}),h_z(\mathbf{k}))$. With this relation between opposite momentum, we can define the unified polar angles and azimuth angles for the two opposite momentum with $\mathbf{h}({\bf k})$. The $m$-derivative of the superconducting phase and the $\bf m$-space electron Berry connection can be calculated straightforwardly, and we find that gauge invariant $m$-derivative of the superconducting phase vanishes
\begin{equation}
    D_{\bf m}\varphi= \nabla_{\bf m}\varphi-\mathfrak{A}^{b}_{n}(\mathbf{k})-\mathfrak{A}^{b}_{n}(-\mathbf{k})=0.
\end{equation}
Similarly, a direct calculation shows that the $m$-derivative of the effective charge for in-plane $\bf m$ also vanishes. As a result, we find that for in-plane $\bf m$ , the first two terms of the Eq.~(\ref{eq:phase-space BC}) vanish, and the phase-space superconducting Berry curvature is the same as the phase-space electron Berry curvature,
\begin{equation}
\Omega_{ k_{\alpha}m_{\beta}}=|u|^2\Omega^{b}_{ k_{\alpha}m_{\beta}}(\mathbf{k})+|v|^2\Omega^{b}_{ k_{\alpha}m_{\beta}}(-\mathbf{k})=\Omega^{b}_{ k_{\alpha}m_{\beta}}(\mathbf{k}),
\end{equation}
where we utilize $\Omega^{b}_{ k_{\alpha}m_{\beta}}(-\mathbf{k})=-\Omega^{b}_{ k_{\alpha}m_{\beta}}(\mathbf{k})$ in the presence of $C_{2z}$ symmetry. The superconducting phase-space Berry curvature for the in-plane $\bf m$ is purely contributed by the electron band, and the superconductivity does not contribute at all. This behavior is entirely different from the momentum-space Berry curvature.

\subsection{Ferromagnetic Rashba model}
We can utilize the above formulas to analytically calculate the Berry curvatures of a two-dimensional toy model where the Ferromagnetic order, the Rashba SOC, and the $s$-wave superconducting gap coexist \cite{sau2010generic}. This model has been widely used in the study of topological superconductivity in proximity-induced two-dimensional superconducting systems \cite{elliott2015colloquium,sato2017topological,aguado2017majorana,lutchyn2018majorana,flensberg2021engineered,yazdani2023hunting,dassarma2023search,tanaka2024theory,amundsen2024}. The electron Hamiltonian is described by Eq.~\eqref{eq:electron H} with
\begin{equation}
   \quad \mathbf{h}=(\alpha_{R} k_y,-\alpha_{R} k_x,V_z),
\end{equation}
where $\alpha_{R}$ denotes the strength of Rashba SOC and $V_z$ is the Zeeman splitting energy. The energy spectrum of the electron is written as $\xi_{\bf k} = \epsilon_{\bf k}-
\mu\pm \sqrt{\alpha_R^2k^2+V_z^2}$ , and the electron Berry curvature is given by $    \Omega_{\bf k}^b= {\alpha_R^2 V_z}/(\alpha_R^2k^2+V_z^2)^{\frac{3}{2}}.$
The proximity to a conventional superconductor would induce an $s$-wave spin-singlet pairing gap $\Delta$ that is a constant in the momentum space. Implementing Eq.~\eqref{eq:2band intra-gap}, we find that the intra-band  effective superconducting gap manifests a chiral $p$-wave form of
\begin{equation}
    \tilde{\Delta}_{\bf k} = \alpha_{R}\Delta (k_y \pm ik_x)/\sqrt{\alpha_R^2k^2+V_z^2}.
\end{equation}
Plugging these expressions into Eq.~\eqref{eq:D k phi for two band model}, we can calculate the gauge invariant $k$-derivative of the superconducting phase as
\begin{equation}
       D_{\bf k}\varphi_{\bf k}=\mp\frac{V_z}{k\sqrt{\alpha_R^2k^2+V_z^2}}\hat{\bf e}_{\zeta},
\end{equation}
with $\hat{e}_\zeta$ denoting the unit polar vector of $\bf h$ in the momentum space, which is parallel to the spin texture induced by the Rashba SOC.
The energy spectrum for the superconducting quasiparticle around the electron Fermi surface is $E_{\bf k } = \sqrt{\xi_{\bf k}^2+|\tilde{\Delta}_{\bf k}|^2} = \sqrt{(h_0\pm h)^2+|\tilde{\Delta}_{\bf k}|^2}$, and the effective charge is given by
\begin{equation}
    \rho_{\bf k}=\frac{\xi_{\bf k}}{E_{\bf k}}=\frac{h_0\pm h}{\sqrt{(h_0\pm h)^2+|\tilde{\Delta}_{\bf k}|^2}}.
\end{equation}
Therefore, the $k$-derivative of effective charge is written as
\begin{equation}
    \nabla_{\mathbf{k}}\rho_{\mathbf{k}}
\approx\frac{h}{\alpha_{R}\Delta k}\left(\nabla_{\mathbf{k}}\epsilon_{{\bf k}}\pm\frac{\alpha_{R}^{2}}{h}\mathbf{k}
\right).
\end{equation}
Plugging the expressions into Eq.~(\ref{eq:Berry curvature for two band model}), we can obtain the superconducting Berry curvature. We can further simplify the result by taking a quadratic dispersion $\epsilon_{\bf k} = k^2/2m$ with $m$ being the effective mass of the electron, and the Berry curvature on the electron Fermi surface is given by
\begin{equation}
        \Omega_{\bf k}=\mp\frac{V_z}{2\alpha_{R}\Delta k}\left(\frac{1}{m}\pm\frac{\alpha_{R}^{2}}{h}\right).
\end{equation}
We note that on the electron Fermi surface, we have $\rho=0$. Therefore, the electron band Berry curvature contribution vanishes.

\section{Model calculation}\label{sec:model}

\begin{figure}[t]
    \centering
    \includegraphics[width=0.9\linewidth]{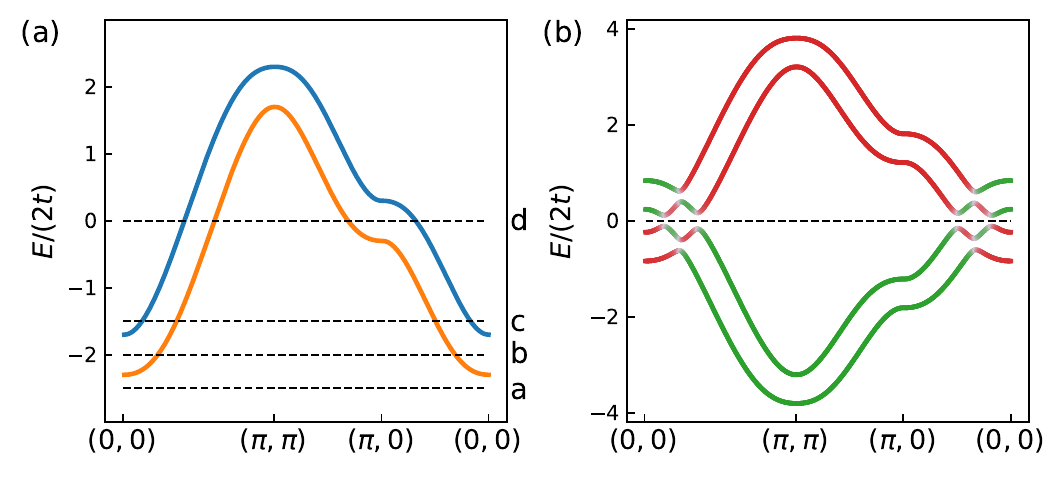}
    \caption{(a) The electron spectrum for the tight-binding ferromagnetic Rashba model. The dashed lines labeled with a-d represent different chemical potentials. (b) Quasiparticle spectrum when the chemical potential cuts both two electron bands as the dashed line c shown in (a). The color denotes the sign of the effective charge of the quasiparticle. The model parameters are taken as $\alpha_R/t=0.2$, $V_z/t=0.15$ and $\Delta/t=0.05$.}
    \label{fig:spectrum}
\end{figure}

We apply the theory developed in the above sections to study the thermo-spin response of the two-dimensional ferromagnetic Rashba model. The tight-binding Hamiltonian of the model on a square lattice is written as \cite{sato2009nonabelian,sau2010generic}
\begin{eqnarray}
    H_{\bf k}&&=[-2t(\cos k_x+\cos k_y)-\mu]\tau_z \\
    +\alpha_R&&(\sin k_y\tau_z\sigma_x-\sin k_x\sigma_y)
    +V_z\tau_z\sigma_z+i\Delta\tau_y\sigma_y,\nonumber
\end{eqnarray}
where $t$ denotes the nearest hopping energy, $\mu$ represents the chemical potential, $\alpha_R$ and $V_z$ are spin splitting energies of the Rashba-SOC and Zeeman field, and $\Delta$ stands for the $s$-wave pairing gap function. This model exhibits rich topological phases labeled by distinct Chern numbers spanning from minus one to positive two. Tuning the chemical potential and the Zeeman field, the topological phase boundaries are analytically given by \cite{sato2009nonabelian}
\begin{equation}
    V_z^2 = (\pm 4t - \mu)^2 + \Delta^2, \quad V_z^2 = \mu^2 + \Delta^2.
\end{equation}
Previous studies concentrate on the topological regime where the Chern number is nonzero. However, the topologically trivial regime might be equally interesting from the perspective of Berry curvature effects.
We consider the scenario that the proximity-induced superconducting gap is the smallest energy scale of the model. Then the topological transition can be qualitatively understood by counting the number of electron Fermi surfaces. Let us fix the energies of the Zeeman coupling and the SOC, and take the chemical potential as the fine-tuned parameter. The electron energy spectrum is demonstrated in Fig.~\ref{fig:spectrum}(a), where two electron bands are fully split by the Zeeman energy and the SOC. We consider four distinct chemical potentials, which are demonstrated as dashed lines. For the chemical potential that intersects both two electron bands (dashed line c), the corresponding BdG quasiparticle spectrum is illustrated in Fig. \ref{fig:spectrum}(b). Clearly, the electron Fermi surface corresponds to the conduction band minimum of the BdG quasiparticle spectrum.

\subsection{Berry curvature}

\begin{figure}[b]
\centering 
\includegraphics[width=0.9\linewidth]{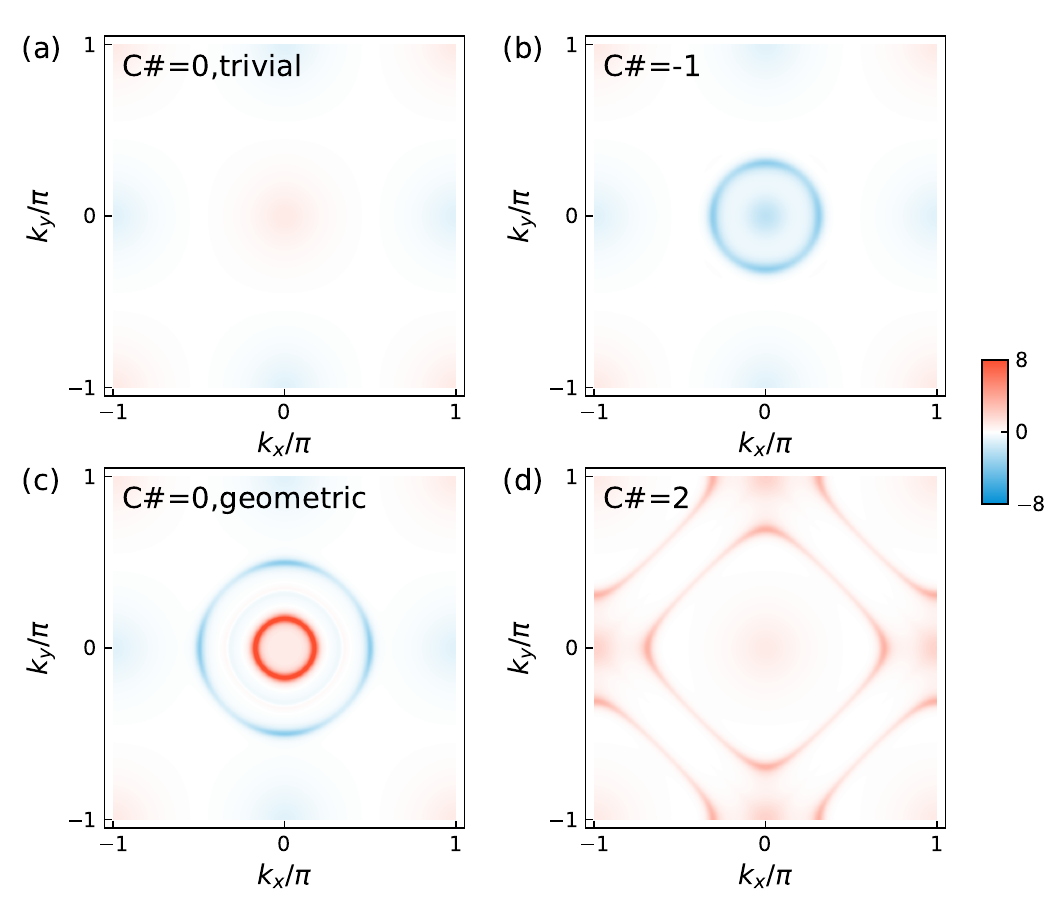}
\caption{Distribution of Momentum-space superconducting Berry curvatures for four typical states with distinct chemical potentials, which are illustrated in Fig.~\ref{fig:spectrum}(a). The Chern numbers are indicated explicitly for each figure.} 
\label{Fig:BcFerro}
\end{figure}

We show the corresponding momentum-space Berry curvature distributions for distinct chemical potentials in Fig.~\ref{Fig:BcFerro} with the Chern number indicated explicitly. The electron system is in the trivial insulating state when the chemical potential is far from the two electron bands (the dashed line a in Fig.~\ref{fig:spectrum}(a)). It is expected that a small superconducting gap will not alter the insulating nature of the electrons. For this case, the quasiparticle Berry curvature is small in the full momentum space, as shown in Fig.~\ref{Fig:BcFerro}(a). The small Berry curvatures around $(0,0)$ and $(0,\pi)$ points are identified as the residual of the electron Berry curvature, which is relatively small when we take a larger Zeeman energy compared with the pairing gap. This phase is trivial not only in the topological sense but also for the geometrical effects of the superconducting quasiparticles, because the superconducting pairing has no contribution to the superconducting Berry curvature. 

When the chemical potential intersects with one of the electron bands (dashed line b in Fig.~\ref{fig:spectrum}(a)), there will be one electron Fermi surface. Then the model enters a topological state with the Chern number equal to minus one. For this case, the distribution of Berry curvature concentrates around the electron Fermi surface, as shown in Fig.~\ref{Fig:BcFerro}(b). In this topological regime, there will be chiral Majorana modes on the boundary of the system. The distributions of the Berry curvatures will induce nontrivial thermo-spin responses that might be useful for detecting these Majorana modes.

If the chemical potential intersects with both electron Fermi bands, there will be two electron Fermi surfaces (dashed lines c and d in Fig.~\ref{fig:spectrum}(a)). There are significant Berry curvatures around the two electron Fermi surfaces as shown in Fig.~\ref{Fig:BcFerro}(c) and Fig.~\ref{Fig:BcFerro}(d). However, the Berry curvatures around the two electron Fermi surfaces may have the same or opposite signs, deciding whether the system is in the topological regime or the trivial regime. The topological regime shown in Fig.~\ref{Fig:BcFerro}(d) somewhat resembles the topological state shown in Fig.~\ref{Fig:BcFerro}(b), but with the results doubled because the Chern number is two. In contrast, the topologically trivial regime must be more interesting because the non-vanishing Berry curvatures at different momentum must cancel. As shown in Fig.~\ref{Fig:BcFerro}(d), there are two closed electron Fermi surfaces possessing the opposite sign Berry curvatures. In this state, the Berry curvatures exactly cancel, and the Chern number is zero. However, the significant Berry curvatures appearing on the electron Fermi surface indicate the non-trivial geometric effects. The induced thermo-spin responses should exhibit intriguing signals even in this geometrical state.


\subsection{Thermo-spin responses}

\begin{figure}[t]
    \centering
    \includegraphics[width=0.95\linewidth]{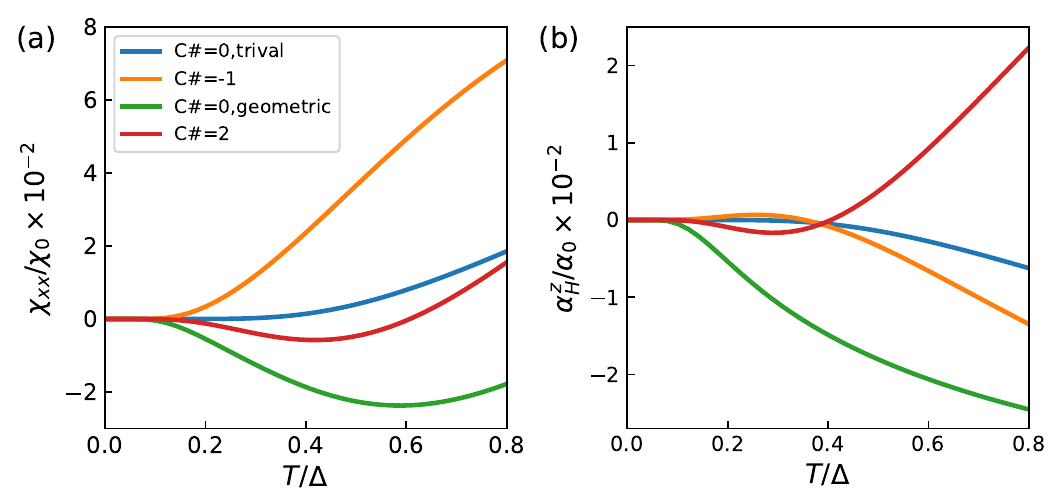}
    \caption{(a) Thermal Edelstein coefficient and (b) spin Nernst conductivity as a function of temperature. The four lines correspond to the four typical model parameters shown in Fig.~\ref{Fig:BcFerro}. Here, $\chi_0=\hbar k_B/2ta$, $\alpha_0=k_B$.}
    \label{fig:enter-label}
\end{figure}

We numerically calculate the Berry curvature induced thermo-spin responses for the four model parameters corresponding to the Berry curvature distributions shown in Fig.~\ref{Fig:BcFerro}. We plot the temperature dependence of the thermal Edelstein coefficient in Fig.~\ref{fig:enter-label}(a). At low temperatures, all the signals are small because the superconducting gap suppresses the excitation of quasiparticles. 
With increasing temperature, the signals are increasing for both topological and geometrical states. In fact, the signal in the topologically trivial state with vanishing Chern number can be larger than the signal of the topological state with Chern number two.
The temperature dependence of spin-Nernst conductivity is shown in Fig.~\ref{fig:enter-label}(b), which is similar to the thermal Edelstein effect.

\section{Summary}
In summary, we derived the semiclassical theory of superconducting quasiparticles for the proximity-induced superconductor in the presence of SOC. We clarify the structure of various Berry connections and Berry curvatures of superconducting quasiparticles, showing the intricate entanglement of pairing and SOC band physics. As applications, we demonstrate the thermal Edelstein effect induced by mixed Berry curvature and the spin Nernst effect induced by momentum space Berry curvature in a model system, illustrating the rich geometric responses in topologically trivial superconducting states.

\appendix

\section{Local pairing Hamiltonian in the band representation}\label{app:local pairing}

Here we give a detailed derivation for the effective gap function in the band representation. We introduce the center of mass coordinate ${\bf R}=\frac{1}{2}({\bf r}_1+{\bf r_2})$ and the relative coordinate ${\bf r}={\bf r}_1-{\bf r}_2$ for the two electrons in a Cooper pair. The local pairing Hamiltonian \eqref{eq:localgap} is thus rewritten as
\begin{equation} \label{eq:localgap Hamiltonian}
\hat{H}^{1}_{c} = \iint d{\bf R} d{\bf r}\Delta^c_{\sigma\sigma'}({\bf r})c_{\sigma}^{\dagger}({\bf R}+\frac{\bf r}{2})c_{\sigma'}^{\dagger}({\bf R}-\frac{\bf r}{2})+h.c.,
\end{equation}
with the local pairing gap function given as
\begin{equation}
    \Delta^c_{\sigma\sigma'}({\bf r})=\Delta({\bf r}_c) \chi_{\sigma\sigma'}({\bf r})e^{i{\bf R} \cdot{\bf \nabla_{{\bf r}_c}\varphi({\bf r}_{\mathbf{c}})}},
\end{equation}
where $\Delta({\bf r}_c) $ denotes the gap function at the center of the wavepacket, $\varphi({\bf r}_{\mathit{c}})$ is the phase of $\Delta({\bf r}_c) $, and $\chi_{\sigma\sigma'}({\bf r})$ represents the relative wave function of two electrons in a Cooper pair. We plug the inverse transformation of Eq.~\eqref{eq:Fourier} into the Eq.~(\ref{eq:localgap Hamiltonian}) and obtain
\begin{eqnarray}\label{eq:app a1}
    \hat{H}^{1}_{c}=&&\Delta({\bf r}_c)\sum_{nn'{\bf k}_1{\bf k}_2}c^{\dagger}_{n{\bf k_1}}c^{\dagger}_{n'{\bf k}_2} \int d{\bf r} \chi_{\sigma\sigma'}({\bf r}) e^{-\frac{i}{2}({\bf k}_1-{\bf k}_2)\cdot {\bf r}}\nonumber\\
    \times\int d{\bf R}&&\phi^*_{n{\bf k}_1,\sigma}({\bf R}+\frac{\bf r}{2})\phi^*_{n'{\bf k}_2,\sigma'}({\bf R}-\frac{\bf r}{2})e^{-i({\bf k}_1+{\bf k}_2-{\bf q})\cdot {\bf R}}+h.c.,\nonumber\\
\end{eqnarray}
where ${\bf q}=\nabla_{{\bf r}_c}\varphi({\bf r}_c)-2{\bf A}({\bf r}_c)$ denotes the Cooper pairing momentum. We can integrate out the plane wave part by exploiting the cell-periodic property of the Bloch wave function, and reduce the integration over the full real space to the integration over one unit cell for ${\bf R}$ as
\begin{eqnarray}\label{eq:integral R}
    &&\int d{\bf R}\phi^*_{n{\bf k}_1,\sigma}({\bf R}+\frac{\bf r}{2})\phi^*_{n'{\bf k}_2,\sigma'}({\bf R}-\frac{\bf r}{2})e^{-i({\bf k}_1+{\bf k}_2-{\bf q})\cdot {\bf R}}\nonumber\\
    =&&\int _{\bf R}\phi^*_{n{\bf k}_1,\sigma}({\bf R}+\frac{\bf r}{2})\phi^*_{n'{\bf k}_2,\sigma'}({\bf R}-\frac{\bf r}{2})
     \delta({\bf k}_1+{\bf k}_2-{\bf q}),
\end{eqnarray}
where $\int_{\bf R}$ denotes the integral over one unit cell.
Implementing the above result into Eq.~\eqref{eq:app a1} and relabeling the momentum subscript as ${\bf k_1}={\bf k}+\frac{\bf q}{2}$ and ${\bf k}_2=-{\bf k}+\frac{\bf q}{2}$, one obtains the local pairing Hamiltonian:
\begin{align}
    \hat{H}^{1}_{c}&=\Delta({\bf r}_c)\sum_{nn'{\bf k}}c^{\dagger}_{n{\bf k}+\frac{\bf q}{2}}c^{\dagger}_{n'-{\bf k}+\frac{\bf q}{2}} \int d{\bf r} \chi_{\sigma\sigma'}({\bf r}) e^{-i{\bf k}\cdot {\bf r}}\nonumber\\
    &\times\int _{\bf R}\phi^*_{n{\bf k}+\frac{\bf q}{2},\sigma}({\bf R}+\frac{\bf r}{2})\phi^*_{n'-{\bf k}+\frac{\bf q}{2},\sigma'}({\bf R}-\frac{\bf r}{2})+h.c.
\end{align}
We can separate the Cooper pair relative wave function into a cell-periodic part $\chi^c_{\sigma\sigma'}({\bf r})$ and an envelope part $\chi^e({\bf r})$, and assume that the envelope function is slowly varying in real space with a Fourier transformation of $\chi^e({\bf r})=\sum_{\bf k'}\chi^e_{\bf k'}e^{i{\bf k'}\cdot{\bf r}}$. Then the local pairing Hamiltonian is written as
\begin{eqnarray}
    \hat{H}^{1}_{c}=&&\Delta({\bf r}_c)\sum_{nn'{\bf k}{\bf k}'}c^{\dagger}_{n{\bf k}+\frac{\bf q}{2}}c^{\dagger}_{n'-{\bf k}+\frac{\bf q}{2}} \int d{\bf r} \chi^c_{\sigma\sigma'}({\bf r})\chi^e_{\bf k'} e^{-i({\bf k}-{\bf k}')\cdot {\bf r}}\nonumber\\
    &&\times\int _{\bf R}\phi^*_{n{\bf k}+\frac{\bf q}{2},\sigma}({\bf R}+\frac{\bf r}{2})\phi^*_{n'-{\bf k}+\frac{\bf q}{2},\sigma'}({\bf R}-\frac{\bf r}{2})+h.c.,
\end{eqnarray}
where the integration over $\bf r$ can also be reduced from the full real space to one unit cell and we can arrive at the second term in Eq.~(\ref{eq:localelectron-momentum2}), 
\begin{eqnarray}
    \hat{H}^{1}_{c}=&&\sum_{nn'{\bf k}}\tilde{\Delta}_{nn'{\bf k}} c^{\dagger}_{n{\bf k}+\frac{\bf q}{2}}c^{\dagger}_{n'-{\bf k}+\frac{\bf q}{2}} +h.c.,
\end{eqnarray}
with the effective gap function in the band representation 
\begin{equation}
    \tilde\Delta_{nn'{\bf k}}=\Delta_{\bf k}\int_{\bf R,r}\phi^*_{n{\bf k}+\frac{\bf q}{2},\sigma}({\bf R}+\frac{\bf r}{2})\chi^c_{\sigma\sigma'}({\bf r})\phi^*_{n'-{\bf k}+\frac{\bf q}{2},\sigma'}({\bf R}-\frac{\bf r}{2}),
\end{equation}
where $\Delta_{\bf k}=\Delta({\bf r}_c)\chi^e_{\bf k}$. This is the Eq.~\eqref{eq:gapfunction} of the main text.

\section{Semiclassical Lagrangian and the Berry connections}\label{app:lagrangian}
Here we give a detailed derivation for the wavepacket averaging over the time-evolution operator in Eq.~(\ref{eq:time evolution}), and obtain the Berry connections in the semiclassical Lagrangian. In the derivation, we will need the charge polarization of the wavepacket
\begin{eqnarray}
    {\bf P}^e&=&\langle{\Psi}_{{\bf k}_c}|\tau_z\hat{\bf r}|{\Psi}_{{\bf k}_c}\rangle\nonumber\\
   & =&\int_{{\bf k},{\bf k'}} a_{\bf k}^*a_{\bf k'}\langle\tilde\psi_{\bf k}|e^{-i{\bf k}\cdot {\bf r}}\tau_z(-i\nabla_{\bf k'}e^{i{\bf k}'\cdot {\bf r}})|\tilde\psi_{\bf k'}\rangle \nonumber\\
   &=&\int_{\bf k} a_{\bf k}^*i\nabla_{\bf k} a_{\bf k}\langle\psi_{\bf k}|\tau_z|\psi_{\bf k}\rangle+\langle\psi_{{\bf k}_c}|\tau_z|i\nabla_{\bf k}\psi_{{\bf k}_c}\rangle\nonumber\\
   &=&({\bf r}_c-\mathcal{A}_{\bf k})\rho+\mathcal{A}^e_{\bf k},
\end{eqnarray}
where $\rho$ is the effective charge of the wavepacket given in Eq.~(\ref{eq:effective charge}), and $\mathcal{A}^e_{\bf k}=\langle\psi_{\bf k}|\tau_z|i\nabla_{\bf k}\psi_{\bf k}\rangle$ is the so-called the charge Berry connection in the main text. The charge polarization of the wavepacket can be expressed as ${\bf P}^{e}=\rho{\bf r}_c+{\bf d}^e$ with $\rho{\bf r}_c$ being interpreted as the displacement of the effective charge, and  
 \begin{equation}
     {\bf d}^e=\langle\Psi_{{\bf k}_c}|\tau_z(\hat{\bf r}-{\bf r}_c)|\Psi_{{\bf k}_c}\rangle=\mathcal{A}^e_{\bf k}-\rho\mathcal{A}_{\bf k}
 \end{equation}
 is identified as the charge dipole moment of the quasiparticle wavepacket \cite{wang2021berry,xiao2021conserved,Hsu2025}.

Now we calculate the wavepacket averaging over the time-evolution operator
\begin{equation}\label{eq:T0}
    \langle\Psi_{{\bf k}_c}|i\frac{d}{dt}|\Psi_{{\bf k}_c}\rangle	=\partial_t \gamma_{{\bf k}_c}
	+i\int _{{\bf k}}a_{{\bf k}}^{*}a_{{\bf k}}\langle\tilde{\psi}_{{\bf k}}|\dot{\tilde{\psi}}_{{\bf k}}\rangle,
\end{equation}
where the first term comes from the time derivative of the constructing function, and the second term is the total time derivative of the quasiparticle wave function. The first term can be expressed as 
\begin{equation}\label{eq:T1}
    \partial_{t}\gamma_{{\bf k}_c}=\dot{\gamma}_{{\bf k}_c}-\nabla_{{\bf k}_c}\gamma_{{\bf k}_c}\cdot\dot{\bf k}_c=\dot{\gamma}_{{\bf k}_c}-({\bf r}_c-\mathcal{A}_{{\bf k}_c})\cdot\dot{\bf k}_c,
\end{equation}
where we utilize the formula of the wavepacket position center \eqref{eq:position} to substitute the $k$-derivative of the constructing phase. The total time derivative of the quasiparticle wave function is separated into the direct time derivative and indirect time derivative through the motion of ${\bf r}_c$ as
\begin{equation}
    i\int _{{\bf k}}a_{{\bf k}}^{*}a_{{\bf k}}\langle\tilde{\psi}_{{\bf k}}|\dot{\tilde{\psi}}_{{\bf k}}\rangle
	=\langle{\psi}_{{\bf k}_c}|i\partial_{t}{\psi}_{{\bf k}_c}\rangle + \dot{{\bf r}}_{c}\cdot\int_{{\bf k}}a_{{\bf k}}^{*}a_{{\bf k}}\langle\tilde{\psi}_{{\bf k}}|i\nabla_{{\bf r}_{c}}|\tilde{\psi}_{{\bf k}}\rangle,\label{eq:T2}
\end{equation}
where we assumed that the external gauge field in time-independent. The indirect time derivative through ${\bf r}_c$ is deduced as
\begin{eqnarray}\label{eq:T3}
    && \int_{{\bf k}}a_{{\bf k}}^{*}a_{{\bf k}}\langle\tilde{\psi}_{{\bf k}}|i\nabla_{{\bf r}_{c}}|\tilde{\psi}_{{\bf k}}\rangle\nonumber\\
    &=&\langle{\psi}_{{\bf k}_c}|i\nabla_{{\bf r}_c}{\psi}_{{\bf k}_c}\rangle-\frac{1}{2}\nabla_{{\bf r}_c}\varphi({\bf r}_c)\langle{\psi}_{{\bf k}_c}|\tau_z|{\psi}_{{\bf k}_c}\rangle\nonumber\\
    &&+\sum_{\alpha}\nabla_{{\bf r}_c}{ A}_{\alpha}({\bf r}_c)\langle{\Psi}_{{\bf k}_c}|\tau_z\hat{ r}_{\alpha}|{\Psi}_{{\bf k}_c}\rangle\nonumber\\
    &=&\langle{\psi}_{{\bf k}_c}|i\nabla_{{\bf r}_c}{\psi}_{{\bf k}_c}\rangle-\frac{1}{2}\rho\nabla_{{\bf r}_c}\varphi({\bf r}_c)+\sum_{\alpha}\nabla_{{\bf r}_c}{ A}_{\alpha}({\bf r}_c){P}^e_{\alpha}\nonumber\\
    &=&\langle\psi_{{\bf k}_c}|i\nabla_{{\bf r}_c}\psi_{{\bf k}_c}\rangle+\frac{1}{2}(-\rho{\bf q}+{\bf B}\times{\bf d}^e),
\end{eqnarray}
where $P^e_{\alpha}$ denotes the $\alpha$ component of the charge polarization vector, and a circular gauge ${\bf A}({\bf r}_c)=\frac{1}{2}{\bf B}\times {\bf r}_c$ has been taken for the last line. 

By combining formulae \eqref{eq:T0}-\eqref{eq:T3} with the wavepacket energy expectation \eqref{eq:energy}, the  semiclassical Lagrangian is thus written in the standard form as
\begin{equation}
    \mathcal{L}=-E_c-\dot{\bf k}_c\cdot{\bf r}_c+\dot{\bf k}_c\cdot\mathcal{A}_{{\bf k}_c}+\dot{\bf r}_c\cdot\mathcal{A}_{{\bf r}_c}+\mathcal{A}_t,
\end{equation}
with the superconducting real-space Berry connection
\begin{equation}
    \mathcal{A}_{{\bf r}_c}=\langle\psi_{{\bf k}_c}|i\nabla_{{\bf r}_c}\psi_{{\bf k}_c}\rangle+\frac{1}{2}(-\rho{\bf q}+{\bf B}\times{\bf d}^e).
\end{equation} 
and the time Berry connection
\begin{equation}
    \mathcal{A}_t=\langle\psi_{{\bf k}_c}|i\partial_t\psi_{{\bf k}_c}\rangle.
\end{equation}
Here, the total time-derivative term in the Lagrangian has been omitted, which will not impact the wavepacket dynamics.

\section{Superconducting Berry curvatures}\label{app:BC}

The Berry curvatures in the main text are expressed as derivatives over the Berry connection for simplicity. Here we present the explicit expressions for the Berry curvatures. The real space and momentum space Berry curvatures are antisymmetric tensors which can be written in the vector form as $\Omega^{\alpha}_{\bf k} = \epsilon_{\alpha \beta \gamma} \Omega_{k_{\beta},k_\gamma}$ and $\Omega^{\alpha}_{\bf r} = \epsilon_{\alpha \beta \gamma} \Omega_{r_{\beta},r_\gamma}$ with $\epsilon_{\alpha \beta \gamma}$ the Levi-Civita symbol. These two Berry curvature vectors are written as
\begin{align}
&{\bf \Omega}_{\bf k}=i\langle\nabla_{\bf k}\psi_{\bf k}|\times|\nabla_{\bf k}\psi_{\bf k}\rangle,\nonumber\\
&{\bf \Omega}_{{\bf r}}=i\langle\nabla_{\bf r}\psi_{\bf k}|\times|\nabla_{\bf r}\psi_{\bf k}\rangle+\frac{1}{2}[\rho{\bf B}-\nabla_{{\bf r}}\rho\times{\bf q}-{\bf B}\times(\nabla_{{\bf r}}\times{\bf d}^{e})].
\end{align}
The momentum-time mixed Berry curvature and the position-time mixed Berry curvature are also vectors that write as
\begin{align}
    &\Omega_{{\bf k}t}=-2\textrm{Im}\langle\nabla_{{\bf k}}\psi_{\bf k}|\partial_{t}\psi_{\bf k}\rangle,\nonumber\\
    &\Omega_{{\bf r_{}}t} =-2\textrm{Im}\langle\nabla_{{\bf r}}\psi_{\bf k}|\partial_{t}\psi_{\bf k}\rangle+\frac{1}{2}[-\partial_t(\rho {\bf q}_{})+{\bf B}\times \partial_t {\bf d}^e].
\end{align}
The position-momentum mixed Berry curvatures can only be written as tensors, which are expressed as
\begin{align}
    &{\Omega}_{k_{\alpha}r_{\beta}}=-2\textrm{Im}\langle\partial_{{ k}_\alpha}\psi_{\bf k}|\partial_{{ r}_\beta}\psi_{\bf k}\rangle-\frac{1}{2}( q_{\beta}\partial_{k_{\alpha}}\rho-\epsilon_{\beta\gamma\eta}B_{\gamma}\partial_{k_{\alpha}}d^e_{\eta}),\nonumber\\
    &\Omega_{r_{\alpha}k_{\beta}}=-2\textrm{Im}\langle\partial_{{ r}_\alpha}\psi_{ k}|\partial_{{k}_\beta}\psi_{\bf k}\rangle+\frac{1}{2}(q_{\alpha}\partial_{k_{\beta}}\rho -\epsilon_{\alpha\gamma\eta}B_{\gamma}\partial_{k_{\beta}}d^e_{\eta}).
\end{align}
In this analysis, we have distinctly isolated the Berry curvatures resulting from the external gauge field from those Berry curvatures that originate from the BdG band.

\bibliography{citations}

\end{document}